\documentclass[a4paper,11pt]{article}
\pdfoutput=1 

\usepackage[utf8x]{inputenc}
\usepackage[T1]{fontenc}
\usepackage{tikz}
\usetikzlibrary{matrix,arrows}
\usepackage{enumitem}
\usepackage{mwe}
\usepackage[colorlinks=true
  ,urlcolor=blue
  ,anchorcolor=blue
  ,citecolor=blue
  ,filecolor=blue
  ,linkcolor=blue
  ,menucolor=blue
  ,linktocpage=true
  ,pdfproducer=medialab
  ,pdfa=true]{hyperref}
\usepackage{amsmath}
\usepackage{amsthm}
\usepackage{amssymb}
\usepackage{amsfonts}
\usepackage{mathrsfs}
\usepackage{bbm}
\usepackage{abstract}
\usepackage{graphicx}
\usepackage{placeins}
\usepackage{subfigure}
\usepackage{nicefrac}
\usepackage{booktabs}
\usepackage{bookmark}
\usepackage{multirow}
\usepackage{dsfont}
\usepackage{longtable}
\usepackage{braket}
\usepackage{tikz}
\usepackage{scalerel}
\usepackage{listings}
\usepackage{color}
\usepackage[all]{xy}
\usepackage{tensor}
\usepackage{faktor}
\usepackage{lipsum}

\newcommand{\invRAdS}{L^{-1}}
\newcommand{\invRAdSsq}{L^{-2}}

\newcommand{\holpar}{\alpha} 

\theoremstyle{definition}
\newtheorem{definition}{Definition}[section] 
 
\theoremstyle{definition}
\newtheorem{remark}[definition]{Remark}

\theoremstyle{definition}

\theoremstyle{plain}

\theoremstyle{plain}

\theoremstyle{plain}

\theoremstyle{plain}

\newcommand{\Exterior}{\mathchoice{{\textstyle\bigwedge}}%
    {{\bigwedge}}%
    {{\textstyle\wedge}}%
    {{\scriptstyle\wedge}}}

\usepackage{jheppub} 

\title{\boldmath Supersymmetric minisuperspace models in self-dual loop quantum cosmology}

\author[1]{K. Eder,}
\author[2]{H. Sahlmann,}

\affiliation[]{Friedrich-Alexander-Universität Erlangen-Nürnberg (FAU),\\
Institute for Quantum Gravity (IQG),\\Staudtstr. 7,D-91058 Erlangen, Germany}

\emailAdd{konstantin.eder@gravity.fau.de}
\emailAdd{hanno.sahlmann@gravity.fau.de}

\abstract{In this paper, we study a class of symmetry reduced models of $\mathcal{N}=1$ supergravity using self-dual variables. It is based on a particular Ansatz for the gravitino field as proposed by D'Eath et al. 
We show that the essential part of the constraint algebra in the classical theory closes. In particular, the (graded) Poisson bracket between the left and right supersymmetry constraint reproduces the Hamiltonian constraint. \\
For the quantum theory, we apply techniques from the manifestly supersymmetric approach to loop quantum supergravity, which yields a graded analog of the holonomy-flux algebra and a natural state space. \\
We implement the remaining constraints in the quantum theory. For a certain subclass of these models, we show explicitly that the (graded) commutator of the supersymmetry constraints exactly reproduces the classical Poisson relations. In particular, the trace of the commutator of left and right supersymmetry constraints reproduces the Hamilton constraint operator. Finally, we consider the dynamics of the theory and compare it to a quantization using standard variables and standard minisuperspace techniques.}

\begin{document} 
\maketitle
\flushbottom

\section{Introduction}
First steps towards an extension of the idea of Ashtekar to study gravity with self-dual variables \cite{Ashtekar:1986yd} to supergravity were taken by Jacobson in his seminal paper \cite{Jacobson:1987cj}. Later, these considerations were extended by Fül\"op to include a cosmological constant \cite{Fulop:1993wi}. Moreover, Fül\"op made the interesting observation that the classical Poisson algebra generated by the Gauss and the left supersymmetry constraint obeys the structure of a graded Lie algebra, leading to a super variant of the self-dual Ashtekar connection. Based on these observations, Gambini et al. \cite{Gambini:1995db} and Ling and Smolin \cite{Ling:1999gn} began to extend the standard quantization techniques of loop quantum gravity to the super category, introducing a notion of super spin network states. A treatment of the theory as a constrained BF theory can be found in \cite{Ling:2000ss}. Later, these ideas were also picked up by Livine and Oeckl \cite{Livine:2003hn} in order to get an idea to quantize fermions in the spin foam approach. In a major step, Bodendorfer, Thiemann and Thurn included higher dimensions and higher supersymmetry, \cite{Bodendorfer:2011pb,Bodendorfer:2011pc,Bodendorfer:2011hs}. These works use real connection variables that allow  to retain a compact structure group. However, the supersymmetry is not manifest anymore in these variables. Ashterkar-like variables were employed by Melosch and Nicolai in a higher dimensional context \cite{Melosch:1997wm}.\\
\\
Based on the ideas of the D'Auria-Fr\'e approach to supergravity \cite{DAuria:1980cmy,DAuria:1982uck,Castellani:1982kd}, a mathematical rigorous geometric approach towards $D=4, \mathcal{N}=1$ supergravity in terms of a supersymmetric generalization of a Cartan connection was considered in \cite{Eder:2020erq}. Moreover, it was shown that due to the structure of the relevant supergroups, the chiral decomposition of the complexified Lorentz group can be extended to the supergroups, and consequently that there is a super Ashtekar connection that can be interpreted as a generalized super Cartan connection and therefore indeed gives chiral supergravity the structure of a super Yang-Mills theory. This, on the one hand, provides a mathematical rigorous foundation for an approach to loop quantum supergravity as studied e.g. in \cite{Fulop:1993wi,Ling:1999gn,Gambini:1995db,Ling:2000ss,Livine:2003hn} that, at least partially, keeps supersymmetry manifest throughout, and, on the other hand, may open the possibility for a systematic study of these type of theories. Furthermore, it was shown that this particular structure of the underlying supersymmetry algebra is even preserved in case of extended supersymmetry. In fact, chiral supergravity theories with extended supersymmetry have been studied in \cite{Sano:1992jw,Ezawa:1995nj,Tsuda:2000er}, where it was shown explicitly in case $\mathcal{N}=2$ that the structure of the chiral supersymmetry algebra is maintained in the canonical theory. This demonstrates that the existence of the graded analog of Ashtekar's connection in $\mathcal{N}=1$ supergravity is not just a mere coincidence. Since extended supersymmetry also involves higher gauge fields \cite{Fiorenza:2013nha,Sati:2015yda}, this may, among other things, provide important insights for quantizing higher gauge fields in the framework of loop quantum gravity.\\
\\
In the present work, we want to study a symmetry reduced model of chiral $\mathcal{N}=1$ supergravity. This will, on the one hand, allow us to point out some interesting structures underlying the manifest approach to supergravity such as the treatment of fermionic fields in the quantum theory. On the other hand, this will provide a first approach to study implications of supersymmetry in the framework of loop quantum cosmology. Moreover, studying self-dual variables is of course an interesting topic in itself and our model will extend previous considerations by Wilson-Ewing  \cite{Wilson-Ewing:2015lia} by including fermions and supersymmetry.
For the quantization of the model using non-chiral variables see \cite{1987CQGra...4.1477M,1993PhLB..300...44D}.\\
For this model, we apply a particular ansatz for the fermion fields as proposed by D'Eath et al. \cite{DEath:1996ejf,DEath:1988jyp,DEath:1992wve}. As we demonstrate in section \ref{sec:symred}, under certain additional requirements concerning the reality conditions, this ansatz can be justified by considering hybrid homogeneous (isotropic) super connection forms as studied more systematically in \cite{Eder:2020erq}. With this ansatz, we  symmetry reduce the chiral action and derive the reduced left and right supersymmetry constraints as well as the Hamiltonian constraint. Moreover, it will be shown 
that the essential part of the constraint algebra in the classical theory closes. In particular, the (graded) Poisson bracket between the left and right supersymmetry constraint reproduces the Hamiltonian constraint modulo the right SUSY constraint. In section \ref{sec:quantumtheory}, we will then go over to the quantum theory and construct the kinematical Hilbert space of super loop quantum cosmology. We will therefore motivate the state space studying the super holonomies of the super Ashtekar connection. In this way, we derive a graded analog of the classical holonomy flux algebra in section \ref{sec:quantumalgebra}. The quantization of this theory is then performed, by considering representations of this algebra on a graded Hilbert space in \ref{sec:rep}. Since these representation are required to be grading preserving, this automatically yields the correct statistics for the bosonic and fermionic degrees of freedom. Finally, we study the implementation of the reality conditions.\\
In section \ref{SUSY constraint} we implement the dynamical constraints in the quantum theory given by the SUSY constraints and Hamiltonian constraint. For a certain subclass of these models, we will show 
that the (graded) commutator of the supersymmetry constraints exactly reproduces the classical Poisson relations. In particular, the trace of the commutator of left and right supersymmetry constraints reproduces the Hamilton constraint. The requirement of this closure fixes some of the quantization ambiguities. 
In section \ref{sec:semi} we study the semiclassical limit of the theory in which quantum corrections arising from quantum geometry are supposed to be negligible. We derive the form of the left and right SUSY constraint in this limit and study their respective solutions. These solutions are then compared to other solutions obtained by different means in the literature. We close with a discussion and outlook in section \ref{sec:disc}. \\
\\
We work in signature $(-+++)$. The gravitational coupling constant is denoted by $\kappa=8 \pi G$, the cosmological constant by $-3/L^2$. Indices $I,J\ldots=0,\ldots,3$ are local Lorentz indices, $i,j,\ldots=1,2,3$ their spatial part. 4D Majorana spinor indices are denoted by $\alpha,\beta,\ldots$ and $A,B,\ldots$ are Weyl spinor indices. The conventions for the spinor calculations can be found in appendix \ref{Spinor}.

\section{Chiral \texorpdfstring{$\mathcal{N}=1$}{TEXT} Supergravity}\label{section:2}
In this section, let us briefly review chiral $\mathcal{N}=1$ (anti-de Sitter) supergravity as introduced in \cite{Jacobson:1987cj} as well as its interpretation in terms of a super Cartan geometry (for more details, we refer to \cite{Eder:2020erq}).\\
Anti-de Sitter supergravity in $D=4$ and $\mathcal{N}=1$ can be described as a super Cartan geometry $(\pi:\,\mathcal{P}\rightarrow\mathcal{M},\mathcal{A})$ modeled on a super Klein geometry $(\mathrm{OSp}(1|4),\mathrm{Spin}^+(1,3))$. Here $\mathrm{OSp}(1|4)$ is an orthosymplectic super Lie group, the super extension of the isometry group of anti-de Sitter space $\mathrm{AdS}_4$ in four spacetime dimensions. The even generators of the corresponding super Lie algebra are given by the translations $P_I$ as well as Lorentz transformations $M_{IJ}$, $I,J=1,\ldots,3$, whereas the odd part of the super Lie algebra is generated by 4 Majorana fermions $Q_{\alpha}$. The super Cartan connection $\mathcal{A}\in\Omega^1(\mathcal{P},\mathfrak{osp}(1|4))$ then decomposes according to
\begin{equation}
\mathcal{A}=e^IP_I+\frac{1}{2}\omega^{IJ}M_{IJ}+\psi^{\alpha}Q_{\alpha}
\end{equation}
with $e^I$ the soldering form (co-frame), $\omega$ the spin connection as well as $\psi$ the Rarita Schwinger field. The symmetry group $\mathrm{OSp}(1|4)$ contains a proper chiral and anti chiral sub super Lie algebra generated by $(T^+_i,Q_A)$ and $(T_i^{-},\bar{Q}^{A'})$, respectively, where
\begin{equation}
T_i^{\pm}:=\frac{1}{2}(J_i\pm iK_i)
\end{equation}     
with $J_i$ and $\tilde{K}_i$ the generators of rotations and boosts, respectively. Restricting to the chiral sector, these satisfy the following graded commutation relations
\begin{align}
[T^+_i,T_j^+]&=\tensor{\epsilon}{_{ij}^k}T_k^+\\
[T_i^+,Q_A]&=Q_{B}\tensor{(\tau_i)}{^B_A}\\
[Q_{A},Q_{B}]&=\frac{1}{L}(\epsilon\sigma^i)_{AB}T_i^+
\end{align}
where $L$ is the so-called anti-de Sitter radius. This is precisely the super Lie algebra corresponding the orthosymplectic super Lie group $\mathrm{OSp}(1|2)_{\mathbb{C}}$, the super extension of the isometry group of $D=2$ anti-de Sitter space. Performing a Inönü-Wigner contraction, i.e. taking the limit $L\rightarrow\infty$, this reproduces the super Poincaré group in $D=2$.\\
Hence, this observation allows one to define the $\mathfrak{osp}(1|2)_{\mathbb{C}}$-valued 1-form
\begin{equation}
\mathcal{A}^+:=A^{+i}T^{+}_i+\psi^AQ_A
\end{equation}
with $A^{+ i}=\Gamma^i-i K^i$ the self-dual Ashtekar connection, which turns out to have the structure of a generalized super Cartan connection. This is precisely the super Asthtekar connection as first introduced in \cite{Fulop:1993wi}. The action of chiral $\mathcal{N}=1$ supergravity takes the form \cite{Jacobson:1987cj} (see appendix \ref{Spinor} for our choice of conventions)
\begin{align}
S(e,\mathcal{A}^+)=\int_M\bigg(&{\frac{i}{\kappa}\tensor{\Sigma}{^{AB}}\wedge\tensor{F(A^+)}{_{AB}}-e_{AA'}\wedge\bar{\psi}^{A'}\wedge D^{(A^+)}\psi^A}\label{eq:0.2.1}\\
&-\frac{1}{2L}\tensor{\Sigma}{^{AB}}\wedge\psi_A\wedge\psi_B+\frac{1}{2L}\tensor{\Sigma}{^{A'B'}}\wedge\bar{\psi}_{A'}\wedge\bar{\psi}_{B'}+\frac{i}{4\kappa L^2}\tensor{\Sigma}{^{AB}}\wedge\tensor{\Sigma}{_{AB}}\bigg)\nonumber
\end{align}
with $\kappa=8\pi G$. Here, $\bar{\psi}^{A'}$ is the complex conjugate of $\psi^A$, $D^{(A^+)}\psi^A=\mathrm{d}\psi^A+\tensor{{A^{+}}}{^A_B}\wedge\psi^B$ denotes the exterior covariant derivative associated to $A^+$ in the fundamental representation of $\mathrm{SL}(2,\mathbb{C})$ acting on purely unprimed indices and $F(A^+)=\mathrm{d}A^{+}+A^{+}\wedge A^+$ is the curvature. Furthermore, $\Sigma^{AA'BB'}:=e^{AA'}\wedge e^{BB'}$ which, due to antisymmetry, can be decomposed according to  
\begin{equation}
\Sigma^{AA'BB'}=\epsilon^{AB}\Sigma^{A'B'}+\epsilon^{A'B'}\Sigma^{AB}
\label{eq:2.2}
\end{equation}
with $\Sigma^{AB}$ and $\Sigma^{A'B'}$ the self-dual anti self-dual part of $\Sigma^{AA'BB'}$, respectively, given by 
\begin{equation}
\Sigma^{AB}:=\frac{1}{2}\epsilon_{A'B'}\Sigma^{AA'BB'}\quad\text{and}\quad\Sigma^{A'B'}:=\frac{1}{2}\epsilon_{AB}\Sigma^{AA'BB'}
\label{eq:2.3}
\end{equation}
Variation of (\ref{eq:0.2.1}) with respect to $\mathcal{A}^+$ implies the field equations of the self dual Ashtekar connection
\begin{equation}
D^{(A^+)}\Sigma^{AB}=-i\kappa\bar{\psi}_{A'}\wedge\psi^{(A}\wedge e^{B)A'}
\label{eq:2.5}
\end{equation}
Taking the complex conjugate of (\ref{eq:2.5}) then yields, provided $e$ is real, the respective equation involving the anti self-dual part $\Sigma^{A'B'}$. Using then identity (\ref{eq:2.2}), this, together with (\ref{eq:2.5}), can be combined to give the respective equation for $\Sigma^{AA'BB'}$ which turns out to be equivalent to 
\begin{equation}
D^{(\omega)}e^{AA'}\equiv\Theta^{(\omega) AA'}=-i\kappa\psi^A\wedge\bar{\psi}^{A'}
\label{eq:2.6}
\end{equation}
with $\Theta^{(\omega) AA'}$ the torsion two-form associated to the spin connection $\omega$ (written in spinor indices). It then immediately follows, provided $e$ is real and (\ref{eq:2.6}) is satisfied, that the action (\ref{eq:0.2.1}) is also purely real (up to a boundary term) \cite{Jacobson:1987cj}. Hence, in this way, one reobtains the ordinary field equations of real $\mathcal{N}=1$ supergravity.\\
\\
The canonical analysis of chiral $\mathcal{N}=1$ supergravity has been investigated in \cite{Jacobson:1987cj,Fulop:1993wi,Ling:2000ss} and, for arbitrary (real) Barbero-Immirzi parameters in \cite{Sawaguchi:2001wi,Tsuda:1999bg,Konsti-SUGRA:2020} and even higher spacetime dimensions in \cite{Bodendorfer:2011pb,Bodendorfer:2011pc}. Here, we will follow \cite{Konsti-SUGRA:2020}, where the results for the 3+1-decomposition of the Holst action are summarized in the appendix \ref{section:Holst}. For $\beta=-i$, it then follows that the chiral action (\ref{eq:0.2.1}) takes the form
\begin{equation}
\begin{split}
S(e,\mathcal{A}^+)&=\int_{\mathbb{R}}{\mathrm{d}t\int_{\Sigma}\mathrm{d}^3x\,\bigg(\frac{i}{\kappa}E^a_i L_{\partial_t}A^{+ i}_a}-\pi^a_A L_{\partial_t}\psi_a^A\\
&\qquad \qquad \qquad\qquad-A_t^i G_i+N^a H_a+N H-S^L_A\psi_t^A-\bar{\psi}_{t A'}S^{R A'}\bigg)
\end{split}\label{eq:2.7}
\end{equation} 
with $\pi_A^a$ the canonically conjugate momentum of $\psi^A_a$ which is related to the corresponding complex conjugate $\bar{\psi}_a^{A'}$ via the \emph{reality condition}
\begin{align}
\pi^a_A&=\tensor{\epsilon}{^{abc}}\bar{\psi}_{b}^{A'}e_{c AA'}
\label{eq:2.8}
\end{align} 
Furthermore, $E^a_i:=|e|e^a_i=\sqrt{q}e^a_i$ is the usual electric field conjugate to $A^{+i}_a$. The canonical pairs $(A^{+i}_a,E^a_i)$ and $(\psi^A_a,\pi_A^a)$ build up a graded symplectic phase space with the nonvanishing Poisson brackets
\begin{equation}
\{E_a^i(x),A_b^{+ j}(y)\}=i\kappa\delta^{j}_i\delta^{(3)}(x,y)\quad\text{and}\quad\{\pi^a_A(x),\psi^B_b(y)\}=-\delta^{a}_b\delta_{A}^B\delta^{(3)}(x,y)
\label{eq:2.9}
\end{equation}
In the decomposition (\ref{eq:2.7}), $S^L_{A}$ and $S^{R A'}$ denote the so-called \emph{left} and \emph{right supersymmetry constraints} which, using (\ref{eq:B2}) as well as the identity $\bar{\psi}_t\gamma_{0i}\psi_a=-2\psi_a^A(\epsilon\tau_i)_{AB}\psi_t^B-2\overline{\psi}_{tA'}(\epsilon\tau_i)^{A'B'}\overline{\psi}_{aB'}$, take the form
\begin{equation}
S^L_A=D_a^{(A^+)}\pi^a_A+\frac{2}{L}E^{ai}\psi_a^B(\epsilon\tau_i)_{BA}
\label{eq:2.10}
\end{equation}
and 
\begin{align}
S^{R A'}=-\epsilon^{A'B'}\epsilon^{abc}\tensor*{e}{_{a AB'}}D_b^{(A^+)}\psi_c^A+\frac{2}{L}E^{ai}\tensor{(\epsilon\tau_i)}{^{A'B'}}\bar{\psi}_{a B'}
\label{eq:2.11}
\end{align}
respectively. According to (\ref{eq:2.11}), the right SUSY constraint depends on the complex conjugate Weyl spinor $\bar{\psi}_{A'}$. In order to re-express in terms of the fundamental variables, one can apply the reality condition (\ref{eq:2.8}). In fact, using (\ref{eq:A1.9})-(\ref{eq:A1.12}), it follows 
\begin{align}
\pi_A^a e_a^{AA'}&=\tensor{\epsilon}{^{abc}}e_a^i\bar{\psi}_{b}^{B'}e^j_c \sigma_i^{AA'}\sigma_{j AB'}\nonumber\\
&=-|e|\epsilon^{ijk}e_k^b\bar{\psi}_{b}^{B'}\sigma_i^{AA'}\sigma_{j AB'}\nonumber\\
&=-2i|e|\bar{\psi}_{b}^{B'}n_{AB'}e^{b AA'}
\end{align}
such that
\begin{align}
-2E^{ai}\tensor{(\epsilon\tau_i)}{^{A'B'}}\bar{\psi}_{a B'}&=2E^{ai}\tensor{(\epsilon\tau_i\epsilon)}{^{A'}_{B'}}\bar{\psi}_{a}^{B'}\nonumber\\
&=-i|e|\bar{\psi}_{a}^{B'}n_{AB'}e^{a AA'}=\frac{1}{2}\pi_A^a e_a^{AA'}
\end{align} 
Together with the identity $\epsilon_{abc}\epsilon^{ijk}E^b_jE^c_k=2\sqrt{q}e^i_a$, we can thus rewrite the right supersymmetry constraint in the equivalent form
\begin{align}
S^{R A'}=-\epsilon^{ijk}\frac{E^b_jE^c_k}{2\sqrt{q}}\sigma_i^{AA'}\left(2\epsilon_{AB}D_{[b}^{(A^+)}\psi_{c]}^B+\frac{1}{2L}\pi^a_A\epsilon_{abc}\right)
\label{eq:2.12}
\end{align}
Finally, the Gauss and vector constraint of the theory are given by
\begin{equation}
G_i=\frac{i}{\kappa}D^{(A^+)}_aE^a_i-\pi^a_A\tensor{(\tau_i)}{^A_B}\psi_a^B
\label{eq:2.13}
\end{equation}
and
\begin{align}
H_a:=\frac{i}{\kappa}E_i^{b}F(A^+)^i_{ab}-\tensor{\epsilon}{^{bcd}}e_{a AA'}\bar{\psi}_{b}^{A'}D^{(A^+)}_{c}\psi_{d}^A
\label{eq:2.14}
\end{align}
respectively. The Hamiltonian constraint reads\footnote{Of course, the vector and Hamiltonian constraint also have to be expressed in terms of the fundamental variables. This can be done in analogy to the right SUSY constraint. We will do so for the symmetry reduced expressions in the following section.}
\begin{align}
H=&-\frac{E_i^aE_j^b}{2\kappa\sqrt{q}}\tensor{\epsilon}{^{ij}_k}F(A^+)^k_{ab}-\epsilon^{abc}\bar{\psi}_a^{A'}n_{AA'}D_b^{(A^+)}\psi_c^A\label{eq:2.15}\\
&+\frac{E^a_iE^b_j}{2L\sqrt{q}}\tensor{\epsilon}{^{ijk}}(\psi_{a A}n_{BA'}\sigma_k^{AA'}\psi_b^B-\bar{\psi}_a^{A'}n_{AA'}\sigma_k^{AB'}\bar{\psi}_{b B'})+\frac{3}{\kappa L^2}\sqrt{q}\nonumber
\end{align}
where $n^{AA'}$ is the spinor corresponding to the unit normal vector field $n^{\mu}$ orthogonal to the time slices $\Sigma_t$ in the 3+1-decomposition.
\begin{remark}\label{remark:2.1}
As an aside, the canonically conjugate momenta $\pi$ and $\tilde{E}:=\frac{i}{\kappa}E$ can be combined to give the super electric field 
\begin{equation}
\mathcal{E}:=\tensor[^i]{T}{^+}\otimes\tilde{E}_i+\tensor[^A]{Q}{}\otimes\pi_A
\label{eq:2.16}
\end{equation}
conjugate to the super connection form $\mathcal{A}^{+}$ where $(\tensor[^i]{T}{^+},\tensor[^A]{Q}{})$ is a so called left-dual basis of $\tensor[^*]{\mathfrak{osp}(1|2)}{_{\mathbb{C}}}$. Based on the fact that $\mathcal{A}^{+}$ is a generalized super Cartan connection, the Gauss and left supersymmetry constraint together generate local $\mathrm{OSp}(1|2)_{\mathbb{C}}$-gauge transformations. As a consequence, both arise from the super Gauss constraint 
\begin{equation}
\mathscr{G}=D^{(\mathcal{A}^+)}_a\mathcal{E}^a
\label{eq:2.17}
\end{equation}
with $D^{(\mathcal{A}^+)}$ the covariant derivative induced by $\mathcal{A}^+$ in the co-adjoint representation.
\end{remark}
Finally, following \cite{Jacobson:1987cj} let us comment on the reality conditions enforced on the self-dual Ashtekar connection in order to recover ordinary real supegravity. As stated above, provided $e$ is real and $A^+$ satisfies the field equation (\ref{eq:2.6}), action (\ref{eq:0.2.1}) is purely real up to a boundary term. Hence, in the canonical theory, it follows that the reality conditions are equivalent to the requirement that the 3D spin connection part $\Gamma^i:=-\tensor{\epsilon}{^i_{jk}}\omega^{jk}$ of $A^+$ satisfies the torsion equation
\begin{equation}
D^{(\Gamma)}e^i\equiv\mathrm{d}e^i+\tensor{\epsilon}{^{i}_{jk}}\Gamma^j\wedge e^k=\Theta^{(\Gamma) i}=\frac{i}{2}\psi^A\wedge\bar{\psi}^{A'}\sigma^i_{AA'}
\label{eq:2.19}
\end{equation} 
which has the unique solution 
\begin{equation}
\Gamma^i\equiv\Gamma^i(e)+C^i(e,\psi,\bar{\psi})
\label{eq:2.20}
\end{equation}
with $\Gamma^i(e)$ the torsion free metric connection
\begin{equation}
\Gamma^i_a(e)=-\epsilon^{ijk}e_j^b\left(\partial_{[a}e_{b]k}+\frac{1}{2}e^c_k e^l_a\partial_{[c}e_{b]l}\right)
\label{eq:2.21}
\end{equation}
and $C^i$ the \emph{contorsion tensor} given by
\begin{equation}
C^i_a=\frac{i\kappa}{4|e|}\epsilon^{bcd}e_d^i\left(2\psi^A_{[a}\bar{\psi}_{b]}^{A'}e_{c AA'}-\psi^A_{b}\bar{\psi}_{c}^{A'}e_{a AA'}\right)
\label{eq:2.22}
\end{equation}
Thus, to summarize, the reality conditions for the bosonic degrees of freedom take the form
\begin{equation}
A^{+ i}_a+(A^{+ i}_a)^*=2\Gamma^i_a(e)+2C^i_a(e,\psi,\bar{\psi}),\quad E^a_i=\Re(E^a_i)
\label{eq:2.23}
\end{equation}
These ensure that, provided the initial conditions satisfy (\ref{eq:2.23}), the dynamical evolution remains in the real sector of the complex phase space, i.e., the phase space of ordinary real $\mathcal{N}=1$ supergravity. 
\section{The symmetry reduced model}
\label{sec:symred}
\subsection{Homogeneous (isotropic) super connection forms}
\label{sec:symredcon}
Typically fermions in cosmological models are not compatible with isotropy. However, in case of $\mathcal{N}=1$ supersymmetry, it turns out that there does exist an ansatz for the gravitino field which is consistent with the requirement of spatial isotropy. This is in fact due to the intrinsic geometric nature of the Rarita-Schwinger field as well as the underlying supersymmetry of the theory and, in the context of loop quantum gravity, can naturally be understood in terms of homogeneous (isotropic) super connection forms which we would like to explain it what follows.\\
The mathemtical theory of invariant super connections is in fact a bit involved as the notion of smoothness in supergeometry is more complicated than in ordinary differential geometry. Moreover, modeling anticommuting classical fermion fields (both with or without presence of supersymmetry) turns out to be by far non-straightforward. However, all this can naturally be resolved combining techniques from algebraic geometry and supergeometry. We will therefore not explain this approach in detail and just sketch the main idea behind it (for more details see \cite{Eder:2020erq}).\\
We consider a spatial slice $\Sigma$ in the spacetime manifold $M$ and assume that $\Sigma$ is homogeneous, i.e., the group $S$ of isometries of $\Sigma$ acts transitively on it. Hence, if $S_x$ denotes the stabilizer or isotropy subgroup at some point $x\in\Sigma$, one can identify $\Sigma$ with the coset space $S/S_x$. Here, in order to compare our results with other results in the literature, we are mainly interested in the standard homogeneous isotropic FLRW models, in particular, in the spatially flat case ($k=0$) and the case with positive spatial curvature $(k=+1)$. In both cases the isometry group takes the form of a semidirect product $S\cong T\rtimes\mathrm{SO}(3)$ with $T$ the subgroup of translations acting freely and transitively on $\Sigma$ and the isotropy subgroup $\mathrm{SO}(3)$.\\
One then asks the question about the existence of a lift of the standard left action of $S$ on $S/S_x$ to left action on the principal $\mathrm{OSp}(1|2)_{\mathbb{C}}$-bundle. Similar to the classical theory, this turns out to be equivalent to classifying conjugacy classes of super Lie group morphisms $\lambda:\,S\rightarrow\mathrm{OSp}(1|2)_{\mathbb{C}}$, where $S$ is viewed as a purely bosonic super Lie group. Given such a super Lie group morphism and the corresponding left action $\hat{f}:\,S\times\mathcal{P}\rightarrow\mathcal{P}$, one can study super connection forms which are invariant w.r.t. the symmetry group, i.e., a super connection $\mathcal{A}^+\in\Omega^1(\mathcal{P},\mathfrak{osp}(1|2)_{\mathbb{C}})_0$ is called \emph{$S$-invariant} if 
\begin{equation}
f_{\phi}^*\mathcal{A}^+=\mathcal{A}^+\quad\forall \phi\in S
\label{eq:3.1}
\end{equation}
It turns out that one actually needs to generalize (\ref{eq:3.1}) in a suitable way such that the connection also encodes (anticommuting) fermionic degrees of freedom. One can then show that any invariant super connection $\mathcal{A}^+$ is uniquely determined by a super Lie algebra morphism $\Lambda:\,\mathrm{Lie}(S)\rightarrow\mathfrak{osp}(1|2)_{\mathbb{C}}$ satisfying $\Lambda|_{\mathrm{Lie}(S_x)}=\lambda_{*}$ and 
\begin{equation}
\Lambda\circ\mathrm{Ad}_{\phi^{-1}}=\mathrm{Ad}_{\lambda(\phi)^{-1}}\circ\Lambda
\label{eq:3.2}
\end{equation}
on $\mathrm{Lie}(S)$ for any $\phi\in S_x$, such that
\begin{equation}
\iota^*\mathcal{A}^+=\Lambda\circ\theta_{\mathrm{MC}}^{(S)}
\label{eq:3.3}
\end{equation}
where $\iota:\,S\rightarrow\mathcal{P}$ is an embedding. Here, $\theta_{\mathrm{MC}}^{(S)}\in\Omega^1(S,\mathrm{Lie}(S))$ is the Maurer-Cartan form on $S$ which satisfies the Maurer-Cartan structure equation
\begin{equation}
\mathrm{d}\theta_{\mathrm{MC}}^{(S)}+\frac{1}{2}[\theta_{\mathrm{MC}}^{(S)}\wedge\theta_{\mathrm{MC}}^{(S)}]=0
\label{eq:3.4}
\end{equation}
In case of FLRW, we have $S\cong T\rtimes\mathrm{SO}(3)$. Let us first assume that $\mathcal{A}^+$ is homogeneous, that is, $\mathcal{A}^+$ is invariant under the translational subgroup $T$ of the full symmetry group. Since $T$ acts freely and transitively on $\Sigma\cong T$, we have $T_x\cong\{e\}$ and the only possible super Lie group morphism $\lambda:\,\{e\}\rightarrow\mathrm{OSp}(1|2)_{\mathbb{C}}$ consists of the identity morphism. Hence, in particular, condition (\ref{eq:3.2}) is empty and it follows that a homogeneous $\mathcal{A}^+$ is uniquely determined by a super Lie algebra morphism $\Lambda:\,\mathrm{Lie}(T)\rightarrow\mathfrak{osp}(1|2)_{\mathbb{C}}$ such that the pullback of $\mathcal{A}^+$ to $\Sigma$ is given by
\begin{equation}
\mathcal{A}^+=\Lambda\circ\theta_{\mathrm{MC}}^{(T)}=:\phi_{i}^kT_{k}^+\mathring{e}^i+\phi^A_{i}Q_A\mathring{e}^i
\label{eq:3.5}
\end{equation} 
where, with respect to a basis $T_i\in\mathrm{Lie}(T)$ of $\mathrm{Lie}(T)$ (not to be confused with chiral generators $T_{i}^+$ of the bosonic subalgebra $\mathfrak{sl}(2,\mathbb{C})$ of $\mathfrak{osp}(1|2)_{\mathbb{C}}$), we set $\phi_i:=\Lambda(T_i)$. As $\mathcal{A}^+$ is even, it follows that $\phi^i$ are bosonic and $\phi^A$ define odd fermion fields. Furthermore, $\mathring{e}^i$ is the induced basis of fiducial left-invariant one-forms\footnote{We are adopting the notations in \cite{DEath:1996ejf,DEath:1992wve} and denote the left-invariant 1-forms by $\mathring{e}^i$ instead of $\mathring{\omega}^i$ as usually done in LQC.} on $\Sigma$. This is the most general form of a homogeneous super connection form.\\
If one requires that $\mathcal{A}^+$, in addition, is isotropic, it then follows that $\Lambda$ needs to satisfy (\ref{eq:3.2}) on the Lie algebra $\mathrm{Lie}(S_x)\cong\mathfrak{so}(3)$ of the isotropy subgroup which infinitesimally reads
\begin{equation}
\Lambda(\mathrm{ad}_{\tau_i}(T_k))=\mathrm{ad}_{\lambda_{*}(T_{i}^+)}(\Lambda(T_k))
\label{eq:3.6}
\end{equation}
$\forall\tau_i\in\mathfrak{so}(3)$ for some super Lie algebra morphism $\lambda_{*}:\,\mathfrak{so}(3)\rightarrow\mathfrak{osp}(1|2)_{\mathbb{C}}$. If one works in the standard category, it follows, as $\lambda_{*}$ is even, that it only takes values in the even Lie subalgebra $\mathfrak{sl}(2,\mathbb{C})$. This corresponds to connections which are invariant under the spatial isotropy group up to gauge transformations. Let us focus first on these type of connections.  We will consider a more general class in remark \ref{Remark:symmetry} below.\\
Since $\mathfrak{so}(3)_{\mathbb{C}}=\mathfrak{sl}(2,\mathbb{C})$, it follows that the only non-trivial Lie algebra morphism, via this identification, is given by the identity morphism, i.e., $\lambda_{*}:\,\mathfrak{so}(3)\rightarrow\mathfrak{sl}(2,\mathbb{C}),\,\tau_i\mapsto T_i^+$. Using that the adjoint representation on the translational subgroup is given by $\mathrm{ad}_{\tau_i}(T_k)=\tensor{\epsilon}{_{ik}^l}T_l$, (\ref{eq:3.6}) leads to the condition
\begin{equation}
\tensor{\epsilon}{_{ik}^l}\phi_l=\mathrm{ad}_{T_i^+}(\phi^j_k T_j^{+}+\phi^A_k Q_A)=\phi^j_k\tensor{\epsilon}{_{ij}^m}T_m^{+}+\phi_k^A\tensor{(\tau_i)}{^B_A}Q_{B}
\label{eq:3.7}
\end{equation}
Restricting first on the even subalgebra, it follows immediately that the unique solution of the bosonic part of the connection has to be of the form 
\begin{equation}
\phi_i^k=c\delta^k_i
\label{eq:3.8}
\end{equation}
for some complex number $c$. This is precisely the form of the isotropic self-dual Ashtekar connection as used in \cite{Wilson-Ewing:2015lia} (see also \cite{Bojowald:2010} for the discussion in case of real variables). Considering next the odd part of the super Lie algebra, one finds that the only possible solution to (\ref{eq:3.7}) for the fermionic degrees of freedom requires $\phi^A_i=0$. Hence, in this restricted subclass of invariant super connections which are invariant only up to ordinary gauge transformations, it follows that one cannot make a purely isotropic ansatz in both bosonic and fermionic degrees of freedom. Nevertheless, as will become clear in what follows, even in this restricted subclass, there does exist a specific ansatz for the homogeneous fermionic fields such that the resulting model is still consistent with spatial isotropy.\\
To this end, following \cite{Bodendorfer:2011pb}, note that the Rarita-Schwinger field $\psi_i:=(\phi^A_i,\bar{\phi}_{i A'})^T$ constructed out of the homogeneous fermionic components of $\mathcal{A}^+$ can be always split into a trace part $\phi:=\gamma^i\psi_i$ as well as a trace-free part $\rho_i:=\psi_i-\frac{1}{3}\gamma_i\phi$ w.r.t. the gamma matrices. Making an isotropic ansatz for the bosonic degrees of freedom, one needs to ensure that the reality condition (\ref{eq:2.23}), which couples bosonic and fermionic degrees of freedom, also becomes isotropic. In particular, the contorsion tensor necessarily has to be of the form
\begin{equation}
C_a^i\equiv C\mathring{e}^i_a
\label{eq:3.9}
\end{equation}
for some real number $C$. As the trace-free part carries internal indices, this will generically not be the case for $\rho_i$. However, as we will see explicitly in section \ref{section:3.2}, the trace part $\phi=\gamma^i\phi_i$ is indeed consistent with the reality condition. For this reason, written again in terms of the chiral components, we make the following ansatz for the odd coefficients of the super Ashtekar connection 
\begin{align}
\phi_i^A&=\sigma_i^{AA'}\bar{\psi}_{A'}\label{eq:3.10a}\\
\bar{\phi}_i^{A'}&=\sigma_i^{AA'}\psi_{A}
\label{eq:3.10b}
\end{align}
This is precisely the ansatz for the Rarita-Schwinger field as proposed by D'Eath at al. in \cite{DEath:1996ejf,DEath:1988jyp,DEath:1992wve} which can be interpreted as a necessary condition on the homogeneous ansatz for the fermionic components of $\mathcal{A}^+$ to provide consistency with the reality condition (\ref{eq:2.23}) in case of purely isotropic bosonic degrees of freedom. 
\begin{remark}
As it was shown in \cite{DEath:1996ejf,DEath:1988jyp,DEath:1992wve}, the ansatz \eqref{eq:3.10a} and \eqref{eq:3.10b} is preserved under a specific subclass of homogeneous supersymmetry transformations. For instance, under a left-handed supersymmetry transformation generated by a homogeneous odd $\epsilon^A$, the fermion field transforms as
\begin{equation}
\delta_{\epsilon}\psi^A_a=D^{(A^+)}_a\epsilon^A=\frac{c}{2i}\tensor{(\sigma_i)}{^{AA'}}n_{BA'}\epsilon^B\mathring{e}^i_a
\label{eq:Variation}
\end{equation}
and thus remains of the form (\ref{eq:3.10a}) if $A^+$ is chosen to be isotropic. Hence, if we take a closer look at the supersymmetry transformation \eqref{eq:Variation}, this suggests that it may be possible to arrive at the ansatz \eqref{eq:3.10a} by considering a more general class of symmetry reduced connections. In fact, for the derivation of \eqref{eq:3.7}, we have assumed that the super Lie algebra morphism $\lambda_*:\mathfrak{so}(3)\rightarrow\mathfrak{osp}(1|2)_{\mathbb{C}}$ only takes values in the subalgebra $\mathfrak{sl}(2,\mathbb{C})$. As a consequence, this describes super connections which are invariant only up to ordinary $\mathrm{SL}(2,\mathbb{C})$-gauge transformations. However, one may also consider super connections which are invariant up to gauge \emph{and} (partial) supersymmetry transformations. To explain this in a bit more detail, let us consider the special case of a vanishing cosmological constant, i.e., $L\rightarrow\infty$ such that, in this limit, $\mathfrak{osp}(1|2)_{\mathbb{C}}$ reduces to the super Poincaré algebra in $D=2$ which we denote by $\overline{\mathfrak{osp}}(1|2)_{\mathbb{C}}$. As shown in \cite{Eder:2020erq}, super connection forms which are invariant up to $\overline{\mathfrak{osp}}(1|2)_{\mathbb{C}}$-super gauge transformations then classified by (even) super Lie algebra morphisms $\lambda_*:\mathfrak{so}(3)\rightarrow\overline{\mathfrak{osp}}(1|2)_{\mathbb{C}}$ which can take values in the whole super Lie algebra. It then follows that a more general class of such morphisms are of the form  
\begin{equation}
    \lambda_{*}^{\psi}(\tau_i)=T_i^++\sigma_i^{AA'}\bar{\psi}_{A'}Q_A
    \label{remark:3.14}
\end{equation}
with some Grassmann-odd $\bar{\psi}_{A'}$. It is then immediate to see that this indeed defines a super Lie algebra morphism, since  
\begin{align}
    [\lambda_{*}^{\psi}(\tau_i),\lambda_{*}^{\psi}(\tau_j)]&=[T_i^+,T_j^+]+\bar{\psi}_{A'}[T_i^+,Q_A]\sigma_j^{AA'}+\bar{\psi}_{A'}[Q_A,T_j^+]\sigma_i^{AA'}\nonumber\\
    &=\tensor{\epsilon}{_{ij}^k} T_k^{+}-i\bar{\psi}_{A'}Q_B\tensor{(\sigma_{[i}\sigma_{j]})}{^{BA'}}\nonumber\\
    &=\tensor{\epsilon}{_{ij}^k}(T_k^++\sigma_k^{AA'}\bar{\psi}_{A'}Q_A)=\tensor{\epsilon}{_{ij}^k}\lambda_{*}^{\psi}(\tau_k)
\end{align}
The form of the reduced connection again follows from the identity \eqref{eq:3.7}. Together with \eqref{remark:3.14}, it follows that the bosonic components of $\mathcal{A}^+$ need to satisfy
\begin{equation}
    \tensor{\epsilon}{_{ik}^l}\phi_l^m=[\lambda_{*}^{\psi}(\tau_i),\phi_k^lT_l^++\phi_k^AQ_A]^m=\phi_k^l\tensor{\epsilon}{_{il}^m}
\end{equation}
and thus again are of the form \eqref{eq:3.8} for some complex $c$. For the fermionic components, this yields 
\begin{align}
    \tensor{\epsilon}{_{ik}^l}\phi_l^A&=[\lambda_{*}^{\psi}(\tau_i),\phi_k^lT_l^++\phi_k^BQ_B]^A=\phi_k^B[T_i^+,Q_B]^A+c\sigma_i^{BB'}\bar{\psi}_{B'}[Q_B,T_k^+]^A\nonumber\\
    &=\phi_k^B\tensor{(\tau_i)}{^A_B}+\frac{c}{2i}\tensor{\epsilon}{_{ik}^l}\sigma_l^{AA'}\bar{\psi}_{A'}
\end{align}
As may be checked by direct computation, this is solved by the ansatz $\phi_i^A=c\sigma_i^{AA'}\bar{\psi}_{A'}$. Thus, we see that allowing for these general type of symmetry reduced connections, this precisely leads to the ansatz of D'Eath. Note, however, that the fermionic fields are not dynamical. This is due to the fact that this consideration only describes connections which are invariant up to left-handed supersymmetry transformations. Hence, for a full treatment, one also has to take right-handed supersymmetry transformations into account. Anyway, this demonstrates that in order to study symmetry reduced models with local supersymmetry, one should consider symmetry reduced connections which are invariant not only up to gauge but, at least in a specific sense, also supersymmetry transformations. 

\end{remark}\label{Remark:symmetry}

\subsection{Symmetry reduction of the chiral action}\label{section:3.2}
Having derived the general ansatz for the bosonic and fermionic degrees of freedom for homogeneous isotropic cosmology, we want to perform a symmetry reduction of the action (\ref{eq:2.7}) and determine the constraints in the symmetry reduced model. Therefore, as already mentioned in the previous section, we are mainly interested in the FLRW models with positive ($k=1$) and vanishing spatial curvature ($k=0$), respectively. In the spatially flat case, the translational subgroup $T$ of the isometry group is Abelian and a basis (co-frame) of fiducial left-invariant one-forms is obtained as the coefficients of the Maurer-Cartan form $\theta^{(T)}_{\mathrm{MC}}=\mathring{e}^i T_i$.\\
In the case $k=+1$, $\Sigma$ is isomorphic to a three sphere $\mathbb{S}^3$ which can be identified with the Lie group $\mathrm{SU}(2)$. Hence, the Maurer-Cartan form of $\mathrm{SU}(2)$ yields a canonical basis of fiducial left-invariant one-forms $\theta^{(\mathrm{SU}(2))}_{\mathrm{MC}}=\mathring{e}^i\tau_i$. For both situations, according to (\ref{eq:3.4}), the structure equation fulfilled by these sets of 1-forms can be written as
\begin{equation}
\mathrm{d}\mathring{e}^i+\frac{k}{2}\tensor{\epsilon}{^i_{jk}}\mathring{e}^j\wedge\mathring{e}^k=0
\label{eq:3.2.1}
\end{equation}
where $k=1$ or $k=0$ in case of a positive or vanishing spatial curvature. The corresponding fiducial frame fields $\mathring{e}_i$ dual to the 1-forms $\mathring{e}^i$ satisfy $\mathring{e}^i_a\mathring{e}^a_j=\delta^i_j$ and form a basis of left-invariant vector fields on $\Sigma$. The fiducial metric $\mathring{q}_{ab}$ is related to the co-frame via
\begin{equation}
\mathring{q}_{ab}=\delta_{ij}\mathring{e}^i_a\mathring{e}^j_b
\label{eq:3.2.2}
\end{equation}
where, for $k=1$, as explained in \cite{Ashtekar:2006es} this metric corresponds to a three-sphere of radius $r_0=2$ so that the total volume of $\Sigma$ is given by $V_0=2\pi^2r_0^3=16\pi^2$. The co-frame $e^i$ of a different spatial slice of the spacetime manifold is related to the fiducial one via rescaling $e^i=a\mathring{e}^i$ and similarly for the triad $e_i=a^{-1}\mathring{e}_i$ where $a$ can be both positive or negative according to the handedness of the triad. Here and in the following, we will fix the sign of the internal three-form $\epsilon_{ijk}$ with the convention $\epsilon_{123}=1$. The volume-form of the spatial slices is then related to the internal 3-from via
\begin{equation}
|e|\epsilon_{abc}=\epsilon_{ijk}e_a^i e_b^j e_c^k
\label{eq:3.2.2.0.1}
\end{equation}
According to (\ref{eq:3.2.2.0.1}), due to the conventions made for the internal 3-form, in case of a positive orientation of the triad, $\epsilon_{abc}$ is normalized to one, i.e., $\epsilon_{123}=+1$. Consequently, changing the orientation of the triad then changes the sign of the volume form such that $\epsilon_{abc}\rightarrow-\epsilon_{abc}$ under $e^i_a\rightarrow-e^i_a$. Finally, from the definition, it follows
\begin{equation}
    \epsilon^{abc}e_a^i e_b^j e_c^k=\epsilon^2\epsilon^{ijk}
\end{equation}
where $\epsilon$ indicates the orientation of the triad. For convenience, following \cite{Wilson-Ewing:2015lia}, we will keep track of the various sign factors appearing in the computations and therefore do not set $\epsilon^2=1$ at this stage, the reason being that this will simplify the implementation of the dynamical constraints in the quantum theory.\\
\\
For the symmetry reduction of the theory, let us fix a fiducial cell $\mathcal{V}$ in $\Sigma$ of finite volume $V_0$ as measured by (\ref{eq:3.2.2}) which will be the whole $\Sigma$ in case $k=1$ or a finite proper sub region in case $k=0$ where, in the latter case, physics will be insensitive to this choice due to homogeneity. Furthermore, we introduce a length scale setting $\ell_0:=V_0^{\frac{1}{3}}$. With these definitions, it follows, using ansatz (\ref{eq:3.8}) for the reduced connection, that the fundamental variables corresponding to the bosonic degrees of freedom take the form
\begin{equation}
 A^{+i}_a=\frac{c}{\ell_0}\mathring{e}^i_a\quad\text{and}\quad E^a_i=\frac{p}{\ell_0^2}\sqrt{\mathring{q}}\mathring{e}^a_i
\label{eq:3.2.3}
\end{equation}
for some real number $p$. For the fermionic degrees of freedom, we choose the ansatz (\ref{eq:3.10a}) and (\ref{eq:3.10b}). The symplectic potential in (\ref{eq:2.7}) then takes the form
\begin{equation}
\int_{\mathbb{R}}\mathrm{d}t\int_{\mathcal{V}}\mathrm{d}^3x\,\left(\frac{i}{\kappa}E^a_i\dot{A}^{+i}_a-\pi^a_A\dot{\psi}^A_a\right)=\int_{\mathbb{R}}\mathrm{d}t\,\left(\frac{3i}{\kappa}p\dot{c}-\bar{\pi}^{A'}\dot{\bar{\psi}}_{A'}\right)
\end{equation}
with the canonically conjugate momentum
\begin{equation}
\bar{\pi}^{A'}:=-6iV_0|a|\psi_An^{AA'}
\label{eq:3.2.4}
\end{equation}
Hence, the canonically conjugate variables in the symmetry reduced theory are given by $(c,p)$ and $(\bar{\psi}_{A'},\bar{\pi}^{A'})$ satisfying the non-vanishing graded Poisson brackets
\begin{equation}
\{p,c\}=\frac{i\kappa}{3}\quad\text{and}\quad\{\bar{\pi}^{A'},\bar{\psi}_{B'}\}=-\delta^{A'}_{B'}
\label{eq:3.2.5}
\end{equation}
where the complex conjugate $\psi_A$ of $\bar{\psi}_{A'}$ is related to its canonically conjugate momentum through the reality condition (\ref{eq:3.2.4}). In order to derrive the respective reality condition in the bosonic sector, let us first consider the metric connection part of (\ref{eq:2.20}) which, using (\ref{eq:3.2.1}), yields
\begin{equation}
\Gamma^i_a(e)=\frac{k}{4}\epsilon^{ijk}\mathring{e}_j^b\left(2\epsilon_{kmn}\mathring{e}^m_{[a}\mathring{e}^n_{b]}+\mathring{e}^c_k \mathring{e}^l_a\epsilon_{lmn}\mathring{e}^m_{[c}\mathring{e}^n_{b]}\right)=\frac{k}{2}\mathring{e}^i_a
\label{eq:3.2.6}
\end{equation} 
Finally, for the torsion contribution we compute, using the identities (\ref{eq:A1.11}) and (\ref{eq:A1.12}) stated in the appendix, 
\begin{align}
C_a^i&=\frac{i\kappa}{4|e|}\epsilon^{bcd}e_d^i\left(2\mathring{e}^{AB'}_{[a}\mathring{e}_{b]}^{BA'}e_{c AA'}-\mathring{e}^{AB'}_{b}\mathring{e}^{BA'}_{c}e_{a AA'}\right)\bar{\psi}_{B'}\psi_B\nonumber\\
&=\frac{i\kappa\epsilon^2}{4 a^2}\epsilon^{ijk}e_a^l\left(\sigma_l^{AB'}\sigma_{j}^{BA'}\sigma_{k AA'}-\sigma_j^{AB'}\sigma_{l}^{BA'}\sigma_{k AA'}-\sigma^{AB'}_{j}\sigma^{BA'}_{k}\sigma_{l AA'}\right)\bar{\psi}_{B'}\psi_B\nonumber\\
&=\frac{i\kappa}{2 a^2}e_a^l\left(-i\sigma_l^{AB'}n_{AA'}\sigma^{i BA'}+\tensor{\epsilon}{^{ij}_l}\sigma_j^{BB'}\right)\bar{\psi}_{B'}\psi_B\nonumber\\
&=\frac{\kappa}{2 a^2}e_a^i n^{BB'}\bar{\psi}_{B'}\psi_B
\label{eq:3.2.7}
\end{align}
which, together with (\ref{eq:3.2.4}), yields
\begin{align}
C_a^i=-\frac{i\kappa}{12\ell_0 p}\mathring{e}_a^i\bar{\pi}^{A'}\bar{\psi}_{A'}
\label{eq:3.2.8}
\end{align}
Hence, the reality condition (\ref{eq:2.23}) for self-dual Ashtekar connection in the reduced theory takes the form
\begin{equation}
c+c^*=k\ell_0-\frac{i\kappa}{6p}\bar{\pi}^{A'}\bar{\psi}_{A'}
\label{eq:3.2.9}
\end{equation}
which, in particular, is isotropic in the torsion contribution consistent with the isotropic ansatz for the gravitational degrees of freedom.\\
\\
Next, we have to compute the constraints in the reduced model. Let us therefore first consider the Gauss constraint (\ref{eq:2.13}). It is immediate to see that the first part depending on the covariant derivative yields a total derivative as the term proportional to the connection simply drops out due to the isotropic ansatz. Hence, only the fermionic contribution remains for which, using (\ref{eq:2.8}), we compute
\begin{align}
G_i&=-\pi^a_A\tensor{(\tau_i)}{^A_B}\psi_a^B=-\epsilon^{abc}\mathring{e}_b^{CA'}\psi_C e_{c AA'}\tensor{(\tau_i)}{^A_B}\mathring{e}_a^{BC'}\bar{\psi}_{C'}\nonumber\\
&=-|a|\sqrt{\mathring{q}}\epsilon^{jkl}\psi_C\bar{\psi}_{C'}\sigma_j^{CA'}\sigma_{k AA'}\tensor{(\tau_i)}{^A_B}\sigma_l^{BC'}\nonumber\\
&=2i|a|\sqrt{\mathring{q}}\psi_C\bar{\psi}_{C'}n_{AA'}\sigma^{l CA'}\tensor{(\tau_i)}{^A_B}\sigma_l^{BC'}\nonumber\\
&=-2i|a|\sqrt{\mathring{q}}\psi_C\bar{\psi}_{C'}n^{CD'}\tensor{(\tau_i)}{_{D'}^{C'}}
\label{eq:3.2.10}
\end{align}
Hence, inserting (\ref{eq:3.2.10}) into the action (\ref{eq:2.7}), we find
\begin{equation}
\int_{\mathbb{R}}\mathrm{d}t\int_{\mathcal{V}}{\mathrm{d}^3x\,A_t^i G_i}=\int_{\mathbb{R}}\mathrm{d}t\,\frac{1}{3}A_t^i\left(-6i|a|V_0\psi_C n^{CD'}\tensor{(\tau_i)}{_{D'}^{C'}}\bar{\psi}_{C'}\right)
\label{eq:3.2.11}
\end{equation}
so that, due to (\ref{eq:3.2.4}), the reduced Gauss constraint can be written in the form
\begin{equation}
G_i=\bar{\pi}^{A'}\tensor{(\tau_i)}{_{A'}^{B'}}\bar{\psi}_{B'}
\label{eq:3.2.12}
\end{equation} 
It is immediate from the homogeneous ansatz that the theory is invariant under diffeomorphism transformations. Hence, concerning the vector constraint, one may expect that (\ref{eq:2.14}) vanishes identically. However, as can be checked explicitly after some lengthy calculation, while the purely bosonic term vanishes identically due the isotropic ansatz for the gravitational degrees of freedom, the fermionic contribution turns out to be proportional to the Gauss constraint. This is in fact analogous to the full theory, where the Gauss constraint needs to be substracted from the vector constraint in order to obtain the infinitesimal generator of pure diffeomorphism transformations.\\
Next, let us turn to the left supersymmetry constraint (\ref{eq:2.10}) which reads
\begin{equation}
S^L_A=\partial_a\pi^a_A-\pi^a_B\tensor{(\tau_i)}{^B_A}A^{+i}_a+2\invRAdS E^{ai}\psi_a^B(\epsilon\tau_i)_{BA}
\label{eq:3.2.13}
\end{equation}
If we drop the total derivative in (\ref{eq:3.2.13}), it follows
\begin{align}
S^L_A&=-\frac{c}{\ell_0}\tensor{\epsilon}{^{abc}}\mathring{e}_a^i\bar{\psi}_{b}^{A'}e_{c BB'}\tensor{(\tau_i)}{^B_A}+2\frac{p}{\ell_0^2L}\sqrt{\mathring{q}}\mathring{e}^{a i}\mathring{e}_a^{BB'}\bar{\psi}_{B'}(\epsilon\tau_i)_{BA}\nonumber\\
&=-\frac{c}{\ell_0 a^2}\epsilon^{abc}e_a^ie_b^je_c^k\psi_C\sigma_j^{CB'}\sigma_{k BB'}\tensor{(\tau_i)}{^B_A}-\frac{i p}{\ell_0^2L}\sqrt{\mathring{q}}\sigma^{i BB'}(\epsilon\tau_i)_{BA}\bar{\psi}_{B'}\nonumber\\
&=\frac{i\epsilon^2 c}{\ell_0 a^2}|a|^3\sqrt{\mathring{q}}\epsilon^{ijk}\tensor{\epsilon}{_{jk}^l}n_{BB'}\sigma_l^{CB'}\tensor{(\tau_i)}{^B_A}\psi_C+\frac{3i  p}{\ell_0^2L}\sqrt{\mathring{q}}\bar{\psi}_{B'}\epsilon^{B'A'}n_{AA'}\nonumber\\
&=\frac{3|a|\epsilon^2}{\ell_0}\sqrt{\mathring{q}} c\psi_A+\frac{3i  p}{\ell_0^2L}\sqrt{\mathring{q}}\bar{\psi}_{B'}\epsilon^{B'A'}n_{AA'}
\label{eq:3.2.14}
\end{align}
which, together with identity (\ref{eq:A1.10}), gives
\begin{equation}
S^L_A=3\sqrt{\mathring{q}}n_{AA'}\left(\frac{|a|\epsilon^2 c}{\ell_0}\psi_Bn^{BA'}+\frac{i p}{\ell_0^2L}\bar{\psi}_{B'}\epsilon^{B'A'}\right)
\label{eq:3.2.15}
\end{equation}
Hence, if we insert the reduced expression (\ref{eq:3.2.14}) into (\ref{eq:2.7}), we find
\begin{align}
\int_{\mathbb{R}}{\mathrm{d}t\int_{\mathcal{V}}{\mathrm{d}^3x\,S^L_A\psi_t^A}}&=\int_{\mathbb{R}}{\mathrm{d}t\,\left(\frac{|a|\epsilon^2 c}{\ell_0}V_0\psi_B n^{BA'}+\frac{i\ell_0}{L} p\bar{\psi}_{B'}\epsilon^{B'A'}\right)3n_{AA'}\psi_t^A}\nonumber\\
&=\int_{\mathbb{R}}{\mathrm{d}t\,\left(-\frac{\epsilon^2 c}{\ell_0}6i|a| V_0\psi_B n^{BA'}+\frac{6\ell_0}{L} p\bar{\psi}_{B'}\epsilon^{B'A'}\right)\frac{i}{2}n_{AA'}\psi_t^A}\nonumber\\
&=:\int_{\mathbb{R}}{\mathrm{d}t\,\left(\frac{\epsilon c}{\ell_0}\bar{\pi}^{A'}+\frac{6\ell_0}{L} |p|\bar{\psi}_{B'}\epsilon^{B'A'}\right)\epsilon\tilde{\psi}_{tA'}}
\label{eq:3.2.16}
\end{align}
where in the last step the reality condition (\ref{eq:3.2.4}) was used. Therefore, in the reduced theory, left supersymmetry constraint takes the form
\begin{equation}
S^{L A'}=\frac{\epsilon c}{\ell_0}\bar{\pi}^{A'}+\frac{6\ell_0}{L} |p|\bar{\psi}_{B'}\epsilon^{B'A'}
\label{eq:3.2.17}
\end{equation}
For the right supersymmetry constraint (\ref{eq:2.11}), we frist focus on the term depending on the covariant derivative which reads
\begin{align}
&-\epsilon^{A'B'}\epsilon^{abc}\tensor*{e}{_{a AB'}}D_b^{(A^+)}\psi_c^A=-\epsilon^{A'B'}\epsilon^{abc}\tensor*{e}{_{a AB'}}(\partial_b\mathring{e}_c^j\sigma_j^{AC'}\bar{\psi}_{C'}+A^{+ j}_b\tensor{(\tau_j)}{^A_C}\psi^C_c)\nonumber\\
=&-\epsilon^{A'B'}\epsilon^{abc}e_a^i\partial_b\mathring{e}_c^j\sigma_{i AB'}\sigma_j^{AC'}\bar{\psi}_{C'}-\epsilon^{A'B'}\epsilon^{abc}\frac{c}{\ell_0}e_a^i\mathring{e}_b^j\mathring{e}_c^k\sigma_{i AB'}\tensor{(\tau_j)}{^A_B}\sigma_k^{BC'}\bar{\psi}_{C'}
\label{eq:3.2.18}
\end{align}
Using again (\ref{eq:3.2.1}), this yields
\begin{align}
&\frac{k}{2}\tensor{\epsilon}{^j_{kl}}\epsilon^{A'B'}\epsilon^{abc}e^i_a\mathring{e}_b^k\mathring{e}_c^l\sigma_{i AB'}\sigma_j^{AC'}\bar{\psi}_{C'}-\frac{c}{\ell_0}\epsilon^{A'B'}\epsilon^2|a|\epsilon^{ijk}\sigma_{i AB'}\tensor{(\tau_j)}{^A_B}\sigma_k^{BC'}\bar{\psi}_{C'}\nonumber\\
=&-3k|a|\epsilon^2\sqrt{\mathring{q}}\epsilon^{A'B'}\bar{\psi}_{B'}+\frac{3|a|\epsilon^2 c}{\ell_0}\sqrt{\mathring{q}}\epsilon^{A'B'}\bar{\psi}_{B'}\nonumber\\
=&\frac{3|a|\epsilon^2}{\ell_0}\sqrt{\mathring{q}}\epsilon^{A'B'}(c-k\ell_0)\bar{\psi}_{B'}
\label{eq:3.2.19}
\end{align}
On the other hand, the term proportional to the cosmological constant gives
\begin{align}
2\invRAdS  E^{ai}\tensor{(\epsilon\tau_i)}{^{A'B'}}\psi_{a B'}&=\frac{2  p}{\ell_0^2L}\sqrt{\mathring{q}}\mathring{e}^{ai}(\epsilon\tau_i)^{A'B'}\mathring{e}_a^{CC'}\epsilon_{C'B'}\psi_C\nonumber\\
&=\frac{2p}{\ell_0^2L}\sqrt{\mathring{q}}\tensor{(\epsilon\tau_i\epsilon)}{^{A'}_{C'}}\sigma^{i CC'}\psi_C\nonumber\\
&=\frac{2p}{\ell_0^2L}\sqrt{\mathring{q}}\tensor{(\tau_i)}{_{C'}^{A'}}\sigma^{i CC'}\psi_C=\frac{3ip}{\ell_0^2L}\sqrt{\mathring{q}}\psi_Cn^{CA'}
\label{eq:3.2.20}
\end{align}
where we used that $\epsilon\tau_i\epsilon=\tau_i^T$. To summarize, we found
\begin{align}
S^{R A'}=\epsilon^{A'B'}\sqrt{\mathring{q}}\left(\frac{3|a|\epsilon^2}{\ell_0}(c-k\ell_0)\bar{\psi}_{B'}+\frac{3i p}{\ell_0^2L}\psi_A n^{AC'}\epsilon_{C'B'}\right)
\label{eq:3.2.21}
\end{align}
Inserting (\ref{eq:3.2.21}) into the action (\ref{eq:2.8}), this yields 
\begin{align}
\int_{\mathbb{R}}{\mathrm{d}t\int_{\mathcal{V}}{\mathrm{d}^3x\,\bar{\psi}_{t A'} S^{R A'}}}&=\int_{\mathbb{R}}{\mathrm{d}t\,-\epsilon\bar{\psi}_{t}^{B'}\left(3a\ell_0^2(c-k\ell_0)\bar{\psi}_{B'}+\frac{3i|p|}{\ell_0^2L} V_0\psi_A n^{AC'}\epsilon_{C'B'}\right)}\nonumber\\
&=\int_{\mathbb{R}}{\mathrm{d}t\,-\epsilon\bar{\psi}_{t}^{B'}\left(3\epsilon\ell_0\sqrt{|p|}(c-k\ell_0)\bar{\psi}_{B'}-\frac{1 }{2L}\sqrt{|p|}\bar{\pi}^{C'}\epsilon_{C'B'}\right)}
\label{eq:3.2.22}
\end{align}
so that, in the reduced theory, the right supersymmetry constraint can be written as
\begin{equation}
S^R_{A'}=3\epsilon\ell_0\sqrt{|p|}(c-k\ell_0)\bar{\psi}_{A'}-\frac{1 }{2L}\sqrt{|p|}\bar{\pi}^{B'}\epsilon_{B'A'}
\label{eq:3.2.23}
\end{equation}
Finally, we need to derive the reduced analog of the Hamiltonian constraint. As far as the purely bosonic term is concerned 
\begin{align}
H_b=-\frac{E_i^aE_j^b}{2\kappa\sqrt{q}}\tensor{\epsilon}{^{ij}_k}F(A^+)^k_{ab}
\label{eq:3.2.24}
\end{align}
in case of the self-dual approach, a reduced expression is stated in \cite{Wilson-Ewing:2015lia}. For sake of completeness, let us derive it here in full detail. This will also clarify the positions of the sign factors. Again applying the structure equation (\ref{eq:3.2.1}), we find for the curvature of the self-dual Ashtekar connection 
\begin{equation}
F(A^+)^k_{ab}=\frac{2c}{\ell_0}\partial_{[a}\mathring{e}_{b]}^k+\frac{c^2}{\ell_0^2}\tensor{\epsilon}{^k_{lm}}\mathring{e}_a^l\mathring{e}_b^m\nonumber\\
=-\frac{kc}{\ell_0}\tensor{\epsilon}{^k_{lm}}\mathring{e}_{a}^l\mathring{e}_b^m+\frac{c^2}{\ell_0^2}\tensor{\epsilon}{^k_{lm}}\mathring{e}_a^l\mathring{e}_b^m
\label{eq:3.2.25}
\end{equation}
such that (\ref{eq:3.2.24}) becomes
\begin{align}
H_b=-\frac{p^2}{2\kappa|a|^3}\sqrt{\mathring{q}}\left(-\frac{6kc}{\ell_0}+\frac{6c^2}{\ell_0^2}\right)=-\frac{3\epsilon^2}{\kappa V_0}\sqrt{|p|}(c^2-k\ell_0 c)\sqrt{\mathring{q}}
\label{eq:3.2.26}
\end{align}
which is exactly the form of the bosonic part of the Hamiltonian constraint as stated in \cite{Wilson-Ewing:2015lia}. Next, let us turn to the fermionic contribution 
\begin{align}
H_f=-\epsilon^{abc}\bar{\psi}_a^{A'}n_{AA'}D_b^{(A^+)}\psi_c^A
\label{eq:3.2.27}
\end{align}
which has in fact a similar form as the right supersymmetry constraint. Following the steps as before, we compute  
\begin{align}
H_f=&-\epsilon^{abc}\bar{\psi}_a^{A'}n_{AA'}D_b^{(A^+)}\psi_c^A=-\epsilon^{abc}\bar{\psi}_a^{B}\mathring{e}_a^{BA'}n_{AA'}(\partial_b\mathring{e}_c^j\sigma_j^{AB'}\bar{\psi}_{B'}+A_b^{+ j}\tensor{(\tau_j)}{^A_C}\mathring{e}_c^k\sigma_k^{CC'}\bar{\psi}_{C'})\nonumber\\
=&\frac{k}{2}\tensor{\epsilon}{^j_{kl}}\epsilon^{abc}\mathring{e}_a^i\mathring{e}_b^k\mathring{e}_c^l(\sigma_i\sigma_j)^{BB'}\psi_B\bar{\psi}_{B'}-\frac{c}{2i\ell_0}\epsilon^{abc}\mathring{e}_a^i\mathring{e}_b^k\mathring{e}_c^l(\sigma_i\sigma_j\sigma_k)^{BB'}\psi_B\bar{\psi}_{B'}\nonumber\\
=&3k\epsilon\sqrt{\mathring{q}}n^{AA'}\psi_A\bar{\psi}_{A'}-\frac{3c}{\ell_0}\epsilon\sqrt{\mathring{q}}n^{AA'}\psi_A\bar{\psi}_{A'}\nonumber\\
=&-\frac{3\epsilon}{\ell_0}\sqrt{\mathring{q}}(c-k\ell_0)n^{AA'}\psi_A\bar{\psi}_{A'}
\label{eq:3.2.28}
\end{align}
Remains to compute the reduced expression of the cosmological contribution in (\ref{eq:2.15}). Some simple algebra reveals 
\begin{align}
H_{\Lambda}=&\sqrt{\mathring{q}}\frac{  p^2}{2\ell_0^4L|a|^3}\tensor{\epsilon}{^{ijk}}\mathring{e}^a_i\mathring{e}^b_j(\psi_{a A}n_{BA'}\sigma_k^{AA'}\psi_b^B-\bar{\psi}_a^{A'}n_{AA'}\sigma_k^{AB'}\bar{\psi}_{b B'})+\sqrt{\mathring{q}}\frac{3 |a|^3}{\kappa L^2}\nonumber\\
=&\sqrt{\mathring{q}}\frac{ \epsilon^2|a|}{2L}\tensor{\mathring{\epsilon}}{^{ijk}}(\sigma_i^{CC'}\epsilon_{CA}n_{BA'}\sigma_k^{AA'}\sigma_j^{BD'}\bar{\psi}_{C'}\bar{\psi}_{D'}-\sigma_i^{CA'}n_{AA'}\sigma_k^{AB'}\sigma_j^{DD'}\epsilon_{D'B'}\psi_{C}\psi_D)\nonumber\\
&+\sqrt{\mathring{q}}\frac{3 |p|^{\frac{3}{2}}}{V_0\kappa L^2}\nonumber\\
=&\sqrt{\mathring{q}}\frac{3i \epsilon^2\sqrt{|p|}}{\ell_0 L}(\epsilon^{C'D'}\bar{\psi}_{C'}\bar{\psi}_{D'}+\epsilon^{CD}\psi_{C}\psi_D)+\sqrt{\mathring{q}}\frac{3|p|^{\frac{3}{2}}}{V_0\kappa L^2}
\label{eq:3.2.30}
\end{align}
Thus, inserting (\ref{eq:3.2.26}), (\ref{eq:3.2.28}) and (\ref{eq:3.2.30}) into (\ref{eq:2.7}), we find
\begin{align}
\int_{\mathbb{R}}{\mathrm{d}t\int_{\mathcal{V}}{\mathrm{d}^3x\,NH}}=&\int_{\mathbb{R}}\mathrm{d}t\,N\left(-\frac{3\epsilon^2}{\kappa }\sqrt{|p|}(c^2-k\ell_0 c)-\frac{\epsilon}{2}(c-k\ell_0)\frac{1}{\sqrt{|p|}}i\bar{\pi}^{A'}\bar{\psi}_{A'}\right.\nonumber\\
&\left.+\frac{3i\epsilon^2\ell_0^2}{L}\sqrt{|p|}(\epsilon^{C'D'}\bar{\psi}_{C'}\bar{\psi}_{D'}+\epsilon^{CD}\psi_{C}\psi_D)+\frac{3|p|^{\frac{3}{2}}}{\kappa L^2}\right)
\label{eq:3.2.31}
\end{align}
Using the reality condition (\ref{eq:3.2.4}), one can express the term in the last line of (\ref{eq:3.2.31}) proportional to the complex conjugate $\psi_A$ of the fundamental variable $\bar{\psi}_{A'}$ in terms of the corresponding canonically conjugate momentum. In fact, direct computation yields
\begin{align}
\epsilon_{A'B'}\bar{\pi}^{A'}\bar{\pi}^{B'}&=-36|a|^2V_0^2n^{AA'}\epsilon_{A'B'}n^{BB'}\psi_A\psi_B\nonumber\\
&=-36|p|\ell_0^4\epsilon^{AB}\psi_A\psi_B
\label{eq:3.2.32}
\end{align}
Hence, it follows that the Hamiltonian constraint in the reduced theory takes the form
\begin{align}
H=&-\frac{3\epsilon^2}{\kappa }\sqrt{|p|}(c^2-k\ell_0 c)-\frac{\epsilon}{2}(c-k\ell_0)\frac{1}{\sqrt{|p|}}i\bar{\pi}^{A'}\bar{\psi}_{A'}\label{eq:3.2.33}\\
&+\frac{3i\epsilon^2\ell_0^2}{L}\sqrt{|p|}\epsilon^{C'D'}\bar{\psi}_{C'}\bar{\psi}_{D'}-\frac{\epsilon^2}{12\ell_0^2 L}\frac{1}{\sqrt{|p|}}\epsilon_{A'B'}\bar{\pi}^{A'}\bar{\pi}^{B'}+\frac{3}{\kappa L^2}|p|^{\frac{3}{2}}\nonumber
\end{align}

\subsection{Half densitized fermion fields}
As proposed in \cite{Thiemann:1997rq} in case of real Ashtekar variables, in order to simplify the reality conditions (\ref{eq:2.8}) for the fermion fields. it is worthwhile to change their density weight going over to half-densities. In the reduced theory, we can do something similar. According to (\ref{eq:3.2.4}), the complex conjugate of $\bar{\psi}_{A'}$ also depends on the scale factor and thus on the momentum conjugate to the reduced connection. Hence, it is suggestive to introduce the new variables\footnote{For notational simplification, we will refrain from indicating the new fundamental variables with an additional bar in what follows. The complex conjugate of $\phi_{A'}$ will then be written as $\bar{\phi}_A$}.
\begin{align}
\phi_{A'}&:=\sqrt{6|a|V_0}\bar{\psi}_{A'}=\sqrt{6}\ell_0|p|^{\frac{1}{4}}\bar{\psi}_{A'}\label{eq:3.3.1}\\
\pi_{\phi}^{A'}&:=\frac{1}{\sqrt{6|a|V_0}}\bar{\pi}^{A'}=\frac{1}{\sqrt{6}\ell_0|p|^{\frac{1}{4}}}\bar{\pi}^{A'}
\label{eq:3.3.2}
\end{align}
Actually, in contrast to the full theory, we are not changing the effective density weight of the fermion fields since, due to the ansatz \eqref{eq:3.10a}, the density weight has already been absorbed in the fiducial co-triad. Nevertheless, these new defined variables have the same dependence on the scale factor $a$ as the half-densitized fields in the full theory (i.e. they are of order $\sim |a|^{\frac{1}{2}}$) which is the reason why we continue to call them half-densities in the reduced theory. Since the definition of the new fields explicitly involves the scale factor (and therefore the reduced electric field $p$), this, a priori, does not provide a canonical transformation. In fact, as will become clear in what follows, this also amounts to a redefinition of the bosonic degrees of freedom.\\
Therefore, let us go back to the symplectic potential (\ref{eq:3.2.4}). Inserting the definitions (\ref{eq:3.3.1}) and (\ref{eq:3.3.2}), we find after some careful analysis  
\begin{align}
\int_{\mathbb{R}}\mathrm{d}t\,\left(\frac{3i}{\kappa}p\dot{c}-\bar{\pi}^{A'}\dot{\bar{\psi}}_{A'}\right)&=\int_{\mathbb{R}}\mathrm{d}t\,\bigg(\frac{3i}{\kappa}p\dot{c}-\sqrt{6|a|V_0}\pi_{\phi}^{A'}L_{\partial_t}\bigg[\frac{1}{\sqrt{6|a|V_0}}\phi_{A'}\bigg]\bigg)\nonumber\\
&=\int_{\mathbb{R}}\mathrm{d}t\,\bigg(\frac{3i}{\kappa}p\dot{c}-\pi_{\phi}^{A'}\dot{\phi}_{A'}+\frac{1}{2}|a|^{-1}\epsilon\dot{a}\pi_{\phi}^{A'}\phi_{A'}\bigg)\nonumber\\
&=\int_{\mathbb{R}}\mathrm{d}t\,\bigg(\frac{3i}{\kappa}p\dot{c}-\pi_{\phi}^{A'}\dot{\phi}_{A'}+\frac{1}{4}\frac{2\epsilon a\dot{a}}{\epsilon a^2}\pi_{\phi}^{A'}\phi_{A'}\bigg)\nonumber\\
&=\int_{\mathbb{R}}\mathrm{d}t\,\bigg(\frac{3i}{\kappa}p\dot{c}-\pi_{\phi}^{A'}\dot{\phi}_{A'}+\frac{1}{4}\frac{\dot{p}}{p}\pi_{\phi}^{A'}\phi_{A'}\bigg)\nonumber\\
&=\int_{\mathbb{R}}\mathrm{d}t\,\bigg(\frac{3i}{\kappa}p L_{\partial_t}\left[c+\frac{i\kappa}{12 p}\pi_{\phi}^{A'}\phi_{A'}\right]-\pi_{\phi}^{A'}\dot{\phi}_{A'}\bigg)
\label{eq:3.3.3}
\end{align}
where in the third line we reinserted the definition $p=\epsilon a^2\ell_0^2$ and, from the fourth to the last line, we integrated by parts dropping a boundary term. Hence, according to (\ref{eq:3.3.3}), this suggests to define the transformation of the reduced connection via
\begin{equation}
\widetilde{c}:=c+\frac{i\kappa}{12 p}\pi_{\phi}^{A'}\phi_{A'}
\label{eq:3.3.4}
\end{equation}
Since signs matter in what follows, it is instructive to check explicitly that this indeed provides a canonical transformation. Observing that $\{c,|p|\}=\{c,\sqrt{p^2}\}=\epsilon\{c,p\}$, direct calculation yields
\begin{align}
\{\widetilde{c},|p|^{\frac{1}{4}}\bar{\psi}_{A'}\}&=\{c+\frac{i\kappa}{12 p}\bar{\pi}^{B'}\bar{\psi}_{B'},|p|^{\frac{1}{4}}\bar{\psi}_{A'}\}\nonumber\\
&=\{c,|p|^{\frac{1}{4}}\}\bar{\psi}_{A'}-\frac{i\kappa\epsilon}{12|p|^{\frac{3}{4}}}\bar{\psi}_{B'}\{\bar{\pi}^{B'},\bar{\psi}_{A'}\}\nonumber\\
&=\frac{1}{4|p|^{\frac{3}{4}}}\epsilon\{c,p\}\bar{\psi}_{A'}+\frac{i\kappa\epsilon}{12|p|^{\frac{3}{4}}}\bar{\psi}_{A'}=0
\label{eq:3.3.5}
\end{align}
proving that the Poisson bracket between $\widetilde{c}$ and $\phi_{A'}$ indeed vanishes. On the other hand, we have
\begin{align}
\{\widetilde{c},|p|^{-\frac{1}{4}}\bar{\pi}^{A'}\}&=\{c,|p|^{-\frac{1}{4}}\}\bar{\pi}^{A'}+\frac{i\kappa\epsilon}{12|p|^{\frac{5}{4}}}\bar{\pi}^{B'}\{\bar{\psi}_{B'},\bar{\pi}^{A'}\}\nonumber\\
&=\frac{i\kappa\epsilon}{12|p|^{\frac{5}{4}}}\bar{\pi}^{A'}-\frac{i\kappa\epsilon}{12|p|^{\frac{5}{4}}}\bar{\pi}^{A'}=0
\label{eq:3.3.6}
\end{align}
and therefore the Poisson bracket between $\widetilde{c}$ and $\pi_{\phi}^{A'}$ is zero, as well. Hence, this proves that the transformation of the phase space variables as declared via (\ref{eq:3.3.1}), (\ref{eq:3.3.2}) as well as (\ref{eq:3.3.4}) indeed provides a canonical transformation.\\
To summarize, w.r.t. to the half densitized fermion fields, the new canonically conjugate variables are given by $(\widetilde{c},p)$ and $(\phi_{A'},\pi^{A'}_{\phi})$, respectively, which satisfy the non-vanishing Poisson brackets 
\begin{equation}
\{p,\widetilde{c}\}=\frac{i\kappa}{3}\quad\text{and}\quad\{\pi_{\phi}^{A'},\phi_{B'}\}=-\delta^{A'}_{B'}
\label{eq:3.3.7}
\end{equation}
The complex conjugate $\bar{\phi}_A$ is related to the canonically conjugate momentum $\pi_{\phi}^{A'}$ via the simplified reality condition
\begin{equation}
\pi_{\phi}^{A'}=-i\bar{\phi}_{A}n^{AA'}\quad\Leftrightarrow\quad\bar{\phi}_A=i\pi_{\phi}^{A'}n_{AA'}
\label{eq:3.3.7.1}
\end{equation}
Using (\ref{eq:3.3.4}), it follows that the respective reality conditions for the bosonic degrees of freedom take the form
\begin{equation}
\widetilde{c}+(\widetilde{c})^{*}=k\ell_0\quad\text{and}\quad p^*=p
\label{eq:3.3.7.2}
\end{equation}
In particular, it follows that, in both parity even odd sector, the torsion contribution in the reality condition of the reduced connection simply drops out! This will drastically simplify their implementation in the quantum theory as will be studied in detail below.\\
Let us rewrite the dynamical constraints in the new fundamental variables. For the Hamiltonian constraint, one immediately finds
\begin{align}
H=&-\frac{3\epsilon^2}{\kappa }\sqrt{|p|}(c^2-k\ell_0 c)-\frac{\epsilon}{2}(c-k\ell_0)\frac{1}{\sqrt{|p|}}i\pi_{\phi}^{A'}\phi_{A'}\label{eq:3.3.8}\\
&+\frac{i \epsilon^2}{2L}\epsilon^{A'B'}\phi_{A'}\phi_{B'}-\frac{i \epsilon^2}{2L}\epsilon_{A'B'}\pi_{\phi}^{A'}\pi_{\phi}^{B'}+\frac{3}{\kappa L^2}|p|^{\frac{3}{2}}\nonumber
\end{align}
where we have implicitly performed the substitution $c:=\widetilde{c}-\frac{i\kappa}{12 |p|}\pi_{\phi}^{A'}\phi_{A'}$. The left and right supersymmetry constraints are given by
\begin{equation}
S^{L A'}=\epsilon c|p|^{\frac{1}{4}}\pi_{\phi}^{A'}+\invRAdS |p|^{\frac{3}{4}}\phi_{B'}\epsilon^{B'A'}
\label{eq:3.3.9}
\end{equation}
and 
\begin{equation}
S^{R}_{A'}=3\epsilon|p|^{\frac{1}{4}}(c-k\ell_0)\phi_{A'}-3\invRAdS |p|^{\frac{3}{4}}\pi_{\phi}^{B'}\epsilon_{B'A'}
\label{eq:3.3.10}
\end{equation}
respectively. 
\begin{remark} 
Some remarks about parity are in order. Due to our sign conventions made for the internal three-form $\epsilon_{ijk}$, the metric connection (\ref{eq:3.2.6}) is even under the parity transformations $e_a^i\rightarrow-e_a^i$  while the extrinsic curvature $K_a^i$ is odd. Hence, $c$ does not have a straightforward behavior under parity transformations. For LQC without fermions another set of conventions, in which the internal three-form $\epsilon_{ijk}$ changes sign under internal parity, is very useful \cite{Ashtekar:2009um}. Under those conventions $c$ does transform in a well defined way and the symplectic structure is preserved under parity transformations. In the present situation, it turns out that those conventions lead to unwanted sign factors in the Dirac algebra (see section \ref{section:Poisson} below), effectively breaking supersymmetry. We suspect that the root of the problem is that the Ashtekar connection has to lift to a connection on both, the frame bundle and the spin structure, in a consistent way. We leave it as a task for future investigation on how to have $c$ transform under parity in a simple way similar to \cite{Ashtekar:2009um} which is also consistent with supersymmetry.\\ 
For completeness, let us also consider the parity transformations of the fermionic fields.
Since the Rarita Schwinger field $\phi:=(\bar{\phi}^{A},\phi_{A'})^T$ , in particular, is a Majorana fermion, one would impose a parity transformation of the form\footnote{Note that, in the mostly plus convention, $(\gamma^0)^2=-\mathds{1}$. Hence, the parity transformation is not involutive but acquires an additional phase of the form $e^{i\frac{\pi}{2}}$. In fact, one cannot simply redefine the parity transformation replacing $\gamma^0$ by $i\gamma^0$ since this turns out to be not compatible with the Majorana condition, i.e., the transformed field will be no longer a Majorana fermion.} $\phi\rightarrow\gamma^0\phi$ which, in terms of the fundamental variables, reads
\begin{equation}
\Pi(\phi)_{A'}:=i\pi_{\phi}^{B'}\epsilon_{B'A'}\quad\text{and}\quad\Pi(\pi_{\phi})^{A'}:=-i\phi_{B'}\epsilon^{B'A'}
\label{eq:3.3.2.1}
\end{equation}
This yields
\begin{equation}
    \Pi(\pi_{\phi}\phi)=\phi_{A'}\pi_{\phi}^{A'}=-\pi_{\phi}\phi
\end{equation}
Hence, it follows that, under these parity transformations, the contorsion tensor is now parity-even. However, in any case, the reduced connection does not seem to have a straightforward behavior under parity transformations. 
\end{remark}
\subsection{Poisson relations}\label{section:Poisson}
We want to study the Poisson relations between the left and the right supersymmtery constraints. Therefore, note that the canonical phase space $\mathscr{P}$ turns out to have the structure of a \emph{super Poisson manifold}. In fact, for any homogenenous supersmooth functions $f,g,h\in\mathcal{O}(\mathscr{P})$ on the phase space, one has
\begin{equation}
\{f,gh\}=\{f,g\}h+(-1)^{|f||g|}g\{f,h\}
\label{eq:3.3.7.3}
\end{equation}
with $|f|\in\mathbb{Z}_2$ called the \emph{parity} of $f$, i.e., $|f|=0$ resp. $|f|=1$ if $f$ is Grassmann even resp. Grassmann odd. That is, according to (\ref{eq:3.3.7.3}), the Poisson bracket defines a super right-derivation on $\mathcal{O}(\mathscr{P})$. Moreover, the Poisson bracket is also graded skew-symmetric, i.e., $\{f,g\}=-(-1)^{|f||g|}\{g,h\}$. Using (\ref{eq:3.3.7.3}), this yields 
\begin{equation}
\{fg,h\}=f\{g,h\}+(-1)^{|g||h|}\{f,h\}g
\label{eq:3.3.7.4}
\end{equation}
and therefore it also defines a super left derivation. In fact, it even follows that the Poisson bracket satisfies a graded analog of the Jacobi identity. Hence, $(\mathcal{O}(\mathscr{P}),\{\cdot,\cdot\})$ has the structure of a super Lie module.\\
With these preparations, let us compute the Poisson bracket between the left and right supersymmetry constraints. One therefore often needs to compute Poisson brackets of the form $\{c,\pi_{\phi}^{A'}\}$ and $\{c,\phi_{A'}\}$ which yields
\begin{equation}
\{c,\pi_{\phi}^{A'}\}=\{\widetilde{c}-\frac{i\kappa}{12p}\pi_{\phi}^{B'}\phi_{B'},\pi_{\phi}^{A'}\}=-\frac{i\kappa}{12p}\pi_{\phi}^{B'}\{\phi_{B'},\pi_{\phi}^{A'}\}=\frac{i\kappa}{12p}\pi_{\phi}^{A'}
\label{eq:3.3.12}
\end{equation}
as well as
\begin{equation}
\{c,\phi_{A'}\}=\{\widetilde{c}-\frac{i\kappa}{12 p}\pi_{\phi}^{B'}\phi_{B'},\phi_{A'}\}=\frac{i\kappa}{12p}\phi_{B'}\{\pi_{\phi}^{B'},\phi_{A'}\}=-\frac{i\kappa}{12p}\phi_{A'}
\label{eq:3.3.13}
\end{equation}
Finally, using 
\begin{equation}
\{\widetilde{c}-\frac{i\kappa}{12 p}\pi_{\phi}^{B'}\phi_{B'},p\}=\{c,p\}=-\frac{i\kappa}{3}
\label{eq:3.3.11}
\end{equation}
as well as the fact that $(\mathcal{O}(\mathscr{P}),\{\cdot,\cdot\})$ defines a super Lie module, it follows for the trace part of the matrix $\{S^{L A'},S^R_{B'}\}$, in case of a vanishing cosmological constant,
\begin{align}
\{\epsilon c|p|^{\frac{1}{4}}\pi_{\phi}^{A'},3\epsilon|p|^{\frac{1}{4}}(c-k\ell_0)\phi_{A'}\}=&-6\epsilon^2\sqrt{|p|}(c^2-k\ell_0 c)+3\epsilon^2\{c,|p|^{\frac{1}{4}}\}(c-k\ell_0)|p|^{\frac{1}{4}}\pi_{\phi}\phi\nonumber\\
&+3\epsilon^2|p|^{\frac{1}{4}}c\{|p|^{\frac{1}{4}},c\}\pi_{\phi}\phi+3\epsilon^2\sqrt{|p|}\pi_{\phi}^{A'}\{c,\phi_{A'}\}(c-k\ell_0)\nonumber\\
&+3\epsilon^2c\sqrt{|p|}\{\pi_{\phi}^{A'},c\}\phi_{A'}\nonumber\\
=&-6\epsilon^2\sqrt{|p|}(c^2-k\ell_0 c)-\frac{i\kappa\epsilon}{4}\frac{(c-k\ell_0)}{\sqrt{|p|}}\pi_{\phi}\phi\nonumber\\
&+\frac{i\kappa\epsilon}{4}\frac{c}{\sqrt{|p|}}\pi_{\phi}\phi-\frac{i\kappa\epsilon}{4}\frac{(c-k\ell_0)}{\sqrt{|p|}}\pi_{\phi}\phi-\frac{i\kappa\epsilon}{4}\frac{c}{\sqrt{|p|}}\pi_{\phi}\phi\nonumber\\
=&2\kappa (H_b+H_f)+\frac{i\kappa\epsilon}{2}\frac{(c-k\ell_0)}{\sqrt{|p|}}\pi_{\phi}\phi\nonumber\\
=&2\kappa H +\frac{i\kappa}{6|p|^{\frac{3}{4}}}\pi_{\phi}^{A'}\left(3\epsilon|p|^{\frac{1}{4}}(c-k\ell_0)\phi_{A'}\right)
\label{eq:3.3.14}
\end{align}
which precisely consists of the sum of the Hamiltonian constraint as well as the right SUSY constraint in case of a vanishing cosmological constant. For $0<L<\infty$, this yields 
\begin{align}
&\{\epsilon c|p|^{\frac{1}{4}}\pi_{\phi}^{A'}+\invRAdS |p|^{\frac{3}{4}}\phi_{B'}\epsilon^{B'A'},3\epsilon|p|^{\frac{1}{4}}(c-k\ell_0)\phi_{A'}-3\invRAdS |p|^{\frac{3}{4}}\pi_{\phi}^{B'}\epsilon_{B'A'}\}\nonumber\\
&=2\kappa (H_{b}+H_f)+\frac{i\kappa}{6|p|^{\frac{3}{4}}}\pi_{\phi}^{A'}\left(3\epsilon|p|^{\frac{1}{4}}(c-k\ell_0)\phi_{A'}\right)\nonumber\\
&\quad+6\invRAdSsq|p|^{\frac{3}{2}}+3\invRAdS \epsilon|p|^{\frac{1}{4}}\{c,|p|^{\frac{3}{4}}\}\pi_{\phi}^{A'}\pi_{\phi}^{B'}\epsilon_{A'B'}-3\invRAdS \epsilon|p|\pi_{\phi}^{A'}\{c,\pi_{\phi}^{B'}\}\epsilon_{B'A'}\nonumber\\
&\quad+3\invRAdS \epsilon|p|^{\frac{1}{4}}\{c,|p|^{\frac{3}{4}}\}\phi_{A'}\phi_{B'}\epsilon^{B'A'}+3\invRAdS \epsilon|p|\phi_{A'}\{c,\phi_{B'}\}\epsilon^{B'A'}\nonumber\\
&=2\kappa (H_{b}+H_f)+\frac{i\kappa}{6|p|^{\frac{3}{4}}}\pi_{\phi}^{A'}\left(3\epsilon|p|^{\frac{1}{4}}(c-k\ell_0)\phi_{A'}\right)\nonumber\\
&\quad+6\invRAdSsq|p|^{\frac{3}{2}}-\frac{3i \kappa}{4L}\epsilon^2\pi_{\phi}^{A'}\pi_{\phi}^{B'}\epsilon_{A'B'}+\frac{i \kappa}{4L}\epsilon^2\pi_{\phi}^{A'}\pi_{\phi}^{B'}\epsilon_{B'A'}\nonumber\\
&\quad+\frac{3i\kappa}{4L}\epsilon^2\phi_{A'}\phi_{B'}\epsilon^{A'B'}+\frac{i \kappa}{4L}\epsilon^2\phi_{A'}\phi_{B'}\epsilon^{A'B'}\nonumber\\
&=2\kappa H-\frac{i\kappa}{2L}\epsilon^2\pi_{\phi}^{A'}\pi_{\phi}^{B'}\epsilon_{A'B'}+\frac{i\kappa}{6|p|^{\frac{3}{4}}}\pi_{\phi}^{A'}\left(3\epsilon|p|^{\frac{1}{4}}(c-k\ell_0)\phi_{A'}\right)\nonumber\\
&=2\kappa H+\frac{i\kappa}{6|p|^{\frac{3}{4}}}\pi_{\phi}^{A'}\left(3\epsilon|p|^{\frac{1}{4}}(c-k\ell_0)\phi_{A'}-3\invRAdS |p|^{\frac{3}{4}}\pi^{B'}\epsilon_{B'A'}\right)
\label{eq:3.3.15}
\end{align}
so that we found
\begin{equation}
\{S^{L A'},S^R_{A'}\}=2\kappa H+\frac{i\kappa}{6|p|^{\frac{3}{4}}}\pi_{\phi}^{A'}S^R_{A'}
\label{eq:3.3.16}
\end{equation}
Let us emphasize that, in the above calculation, the sign factors $\epsilon$ depending on the orientation of the triad matched up exactly to give the algebra (\ref{eq:3.3.16}) proving that the theory is indeed consistent with local supersymmetry in both sectors. Next, we want to compute the off-diagonal entries of the matrix $\{S^{L A'},S^R_{B'}\}$. Following the same steps as before, we immediately find for $A'\neq B'$
\begin{align}
\{\epsilon c|p|^{\frac{1}{4}}\pi_{\phi}^{A'},3\epsilon|p|^{\frac{1}{4}}(c-k\ell_0)\phi_{B'}\}=&-\frac{i\kappa\epsilon}{2}\frac{(c-k\ell_0)}{\sqrt{|p|}}\pi_{\phi}^{A'}\phi_{B'}
\label{eq:3.3.17}
\end{align}
By anticommutativity of fermionic fields, one has $\pi_{\phi}^{A'}\pi_{\phi}^{C'}\epsilon_{C'B'}=0$ for $A'\neq B'$ and similarly for $\phi$. Hence, this yields
\begin{align}
&\{\epsilon c|p|^{\frac{1}{4}}\pi_{\phi}^{A'}+\invRAdS |p|^{\frac{3}{4}}\phi_{C'}\epsilon^{C'A'},3\epsilon|p|^{\frac{1}{4}}(c-k\ell_0)\phi_{B'}-3\invRAdS |p|^{\frac{3}{4}}\pi_{\phi}^{D'}\epsilon_{D'B'}\}\nonumber\\
&=-\frac{i\kappa\epsilon}{2}\frac{(c-k\ell_0)}{\sqrt{|p|}}\pi_{\phi}^{A'}\phi_{B'}=-\frac{i\kappa}{6|p|^{\frac{3}{4}}}\pi_{\phi}^{A'}\left(3\epsilon|p|^{\frac{1}{4}}(c-k\ell_0)\phi_{B'}-3\invRAdS |p|^{\frac{3}{4}}\pi^{C'}\epsilon_{C'B'}\right)
\label{eq:3.3.18}
\end{align}
such that
\begin{equation}
\{S^{L A'},S^R_{B'}\}=-\frac{i\kappa}{6|p|^{\frac{3}{4}}}\pi_{\phi}^{A'}S^R_{B'}
\label{eq:3.3.20}
\end{equation}
Thus, to summarize, we found that
\begin{equation}
\{S^{L A'},S^R_{B'}\}=\left(\kappa H+\frac{i\kappa}{6|p|^{\frac{3}{4}}}\pi_{\phi}^{A'}S^R_{A'}\right)\delta^{A'}_{B'}-\frac{i\kappa}{6|p|^{\frac{3}{4}}}\pi_{\phi}^{A'}S^R_{B'}
\label{eq:3.3.21}
\end{equation}
Equation (\ref{eq:3.3.21}) provides a very strong relation between the Hamiltonian and supersymmetry constraint which will play a central role in section \ref{SUSY constraint} in the construction of the physical sector of the kinematical Hilbert space and the study of the resulting dynamics of the theory.
\section{Quantum theory}
\label{sec:quantumtheory}
\subsection{Construction of the classical algebra}
\label{sec:quantumalgebra}
We want to motivate the kinematical Hilbert space in the reduced theory. We therefore follow the standard procedure in LQC and compute holonomies along straight edges of the fiducial cell $\mathcal{V}$ which are parallel to integral flows $\Phi_{\tau}^{i}$ generated by the basis of left-invariant vector fields $(\mathring{e}_i)_i$ dual to the fiducial co-frame. Choosing the new variables as constructed in the previous section, we combine them to the super connection $\tilde{\mathcal{A}}^+$ defined as  
\begin{equation}
\tilde{\mathcal{A}}^{+}=\tilde{A}^{+}+\tilde{\psi}:=\frac{\tilde{c}}{\ell_0}\mathring{e}^i T_{i}^{+}+\ell_0^{-\frac{3}{2}}\phi_{A'}\mathring{e}^{AA'}Q_{A}
\label{eq:4.1.1}
\end{equation}
where, in the second term, the factor $\ell_0^{-\frac{3}{2}}$ has been included for dimensional reasons. By definition, it follows
\begin{equation}
\{\tilde{\mathcal{A}}^{+\underline{A}}_a,\tilde{\mathcal{A}}^{+\underline{B}}_b\}=0
\end{equation}
for $\underline{A},\underline{B}\in\{i,A\}$. Since, in the manifest approach to canoncial loop quantum supergravity, we have this super connection at our disposal, we would like to motivate the kinematical Hilbert space of the reduced theory studying the corresponding (super) holonomies. These holonomies $h_{e}[\tilde{\mathcal{A}}^+]$ along edges $e\subset\Sigma$ embedded in $\Sigma$ (in case of super matrix Lie groups such as, in the present situation, $\mathrm{OSp}(1|2)_{\mathbb{C}}$) satisfy the equation \cite{Eder:2020erq}
\begin{equation}
\partial_{\tau}h_{e}[\tilde{\mathcal{A}}^+]=-\holpar  \tilde{\mathcal{A}}^{+ e}h_{e}[\tilde{\mathcal{A}}^+]
\label{eq:4.1.2}
\end{equation}
with $\tilde{\mathcal{A}}^{+ e}(\tau):=e^*\tilde{\mathcal{A}}^{+}(\tau)=\dot{e}^a(\tau)\tilde{\mathcal{A}}^{+}_a(e(\tau))$ the pullback of $\tilde{\mathcal{A}}^+$ w.r.t. $e$ and $\holpar\in\mathbb{C}$ some complex number. For a standard holonomy corresponding to a proper parallel transport map one has $\holpar=1$. But, following \cite{Wilson-Ewing:2015lia}, in view of the solution of the reality conditions in the quantum theory (and thus regain the solutions of ordinary real $\mathcal{N}=1$ supergravity) we do not fix this constant to a specific value at this stage. Adopting the terminology of \cite{Wilson-Ewing:2015lia}, we will call them \emph{generalized holonomies}. As shown in \cite{Eder:2020erq}, in a specific gauge, one can decompose the generalized holonomy in the form $h_e[\tilde{\mathcal{A}}^+]=h_e[\tilde{A}^+]\cdot h_e[\tilde{\psi}]$ with $h_{e}[\tilde{A}^+]$ the generalized holonomy generated by the bosonic part $\tilde{A}^+$ of the super connection, such that equation (\ref{eq:4.1.2}) turns out to be equivalent to 
\begin{align}
\partial_{\tau}h_{e}[\tilde{A}^+]&=-\holpar \tilde{A}^{+ e}h_{e}[\tilde{A}^+]\label{eq:4.1.3}\\
\partial_{\tau}h_{e}[\tilde{\psi}]&=-\holpar(\mathrm{Ad}_{h[\tilde{A}^+]^{-1}}\tilde{\psi}^e) h_{e}[\tilde{\psi}]
\label{eq:4.1.4}
\end{align}
where, for $g\in\mathrm{SL}(2,\mathbb{C})$, the adjoint representation on the odd part of the super Lie algebra $\mathfrak{osp}(1|2)_{\mathbb{C}}$ is given by the fundamental representation of $\mathrm{SL}(2,\mathbb{C})$ such that
\begin{equation}
\mathrm{Ad}_{g}\tilde{\psi}=\tilde{\psi}^A Q_{B}\tensor{g}{^B_A}=(\tensor{g}{^B_A}\tilde{\psi}^A)Q_{B}
\label{eq:4.1.5}
\end{equation}
Let us consider edges $\gamma_i:\,[0,b]\rightarrow\Sigma,\,\tau\mapsto\gamma_i(\tau)$ that are parallel along integral flows $\Phi_{\tau}^{i}$ generated by the basis of left-invariant vector fields $(\mathring{e}_i)_i$. We choose $\tau$ as the proper time (length) as measured w.r.t. fiducial metric $\mathring{q}$ on $\Sigma$. Hence, we can identify $b=\ell_0$. With respect to $\gamma_i$, it immediately follows from (\ref{eq:4.1.3}) that $h_{i}[\tilde{A}^+]\equiv h_{\gamma_i}[\tilde{A}^+]$ is given by 
\begin{equation}
h_{i}[\tilde{A}^+](\tau)=\exp\left(-\frac{\holpar\tilde{c}\tau}{\ell_0}T_i^+\right)=\mathrm{cosh}\left(\frac{\holpar\mathring{\mu}\tilde{c}}{2i}\right)\mathds{1}+\mathrm{sinh}\left(\frac{\holpar\mathring{\mu}\tilde{c}}{2i}\right)2iT_i^+
\label{eq:4.1.6}
\end{equation}
where we used that in the fundamental representation of $\mathrm{OSp}(1|2)_{\mathbb{C}}$ one has $(T_i^+)^2=-\frac{1}{4}\mathds{1}$ and set $\mathring{\mu}:=\frac{\tau}{\ell_0}$. Inserting (\ref{eq:4.1.6}) into (\ref{eq:4.1.4}) it follows that the solution of this equation is given by a path ordered exponential yielding
\begin{align}
h_{i}[\tilde{\psi}](\tau)=&\mathcal{P}\exp\left(-\int_0^{\tau}\mathrm{d}\tau'\,\holpar\tensor{\left(e^{\frac{\holpar\tilde{c}\tau'}{\ell_0}T_i^+}\right)}{^B_A}\phi_{A'}\,\sigma^{AA'}_i Q_{B}\right)\nonumber\\
=&\mathds{1}-\int_0^{\tau}\mathrm{d}\tau'\,\holpar\ell_0^{-\frac{3}{2}}\tensor{\left(e^{\frac{\holpar\tilde{c}\tau'}{\ell_0}T_i^+}\right)}{^B_A}\phi_{A'}\,\sigma^{AA'}_i Q_{B}\nonumber\\
&+\int_0^{\tau}\mathrm{d}\tau'\int_0^{\tau'}\mathrm{d}\tau''\holpar^2\ell_0^{-3}\tensor{\left(e^{\frac{\holpar\tilde{c}\tau'}{\ell_0}T_i^+}\right)}{^D_C}\tensor{\left(e^{\frac{\holpar\tilde{c}\tau''}{\ell_0}T_i^+}\right)}{^B_A}\phi_{C'}\phi_{A'}\,\sigma^{CC'}_i\sigma^{AA'}_i  Q_D Q_{B}
\label{eq:4.1.7}
\end{align}
where the sum terminates at second order due to the homogeneity and nilpotency of the fermionic variables. To see how a typical matrix element in (\ref{eq:4.1.7}) looks like, let us compute the first integral which gives
\begin{equation}
\int_0^{\tau}\mathrm{d}\tau'\,\holpar e^{\frac{\holpar\tilde{c}\tau'}{\ell_0}T_i^+}=\frac{4\ell_0}{\tilde{c}}\left(e^{\holpar\tilde{c}\mathring{\mu}T_i^+}-\mathds{1}\right)T_i^+
\label{eq:4.1.8}
\end{equation}
and thus contains terms of the form (\ref{eq:4.1.6}) as well as smooth functions on $\mathbb{C}$ vanishing at infinity. If we first consider the purely bosonic contributions to the generalized super holonomy $h[\tilde{\mathcal{A}}^+]$, it follows that the matrix elements can be equivalently be encoded in terms of holomorphic  functions $f$ on $\mathbb{C}$ of the form $f(z)=e^{Cz}$ with $C\in\mathbb{C}$. In fact, functions of this kind play a special role in group theory. Therefore, consider $(\mathbb{C},+)$ as an additive group. A \emph{generalized character} $\rho$ on $(\mathbb{C},+)$ is defined as a group morphism  
\begin{equation}
\rho:\,(\mathbb{C},+)\rightarrow\mathrm{GL}(1,\mathbb{C})\cong\mathbb{C}^{\times}
\label{eq:4.1.9}
\end{equation}
This implies that $\rho(z+z')=\rho(z)\rho(z')$ for any $z,z'\in\mathbb{C}$ and therefore
\begin{equation}
\rho(z+z')-\rho(z)=\rho(z)\rho(z')-\rho(z)=(\rho(z')-1)\rho(z)
\label{eq:4.1.10}
\end{equation}
Hence, if assume that $\rho$ is differentiable and set $C:=\rho'(0)\in\mathbb{C}$, it immediately follows from (\ref{eq:4.1.10}) 
\begin{equation}
\frac{\mathrm{d}}{\mathrm{d}z}\rho(z)=C\rho(z)
\label{eq:4.1.11}
\end{equation}
which, due to $\rho(1)=e$, has the unique solution
\begin{equation}
\rho(z)=e^{Cz},\quad C\in\mathbb{C}
\label{eq:4.1.12}
\end{equation}
That is, the matrix elements of the bosonic part of the super holonomies can be described in terms of generalized characters on $\mathbb{C}$ as an additive group.
\begin{remark}
This is, in fact, in complete analogy to LQC with real variables. There, it turns out that the matrix elements of holonomies can be encoded in terms ordinary characters of $\mathbb{R}$ regarded as an additive group, i.e. group morphisms
\begin{equation}
\rho:\,(\mathbb{R},+)\rightarrow\mathrm{U}(1)\subset\mathrm{GL}(1,\mathbb{C})
\label{eq:4.1.13}
\end{equation}
The real line $(\mathbb{R},+)$ is the universal covering of $\mathrm{U}(1)$ which is compact. Taking the complexification on both sides of \eqref{eq:4.1.13}, this immediately leads to the notion of a generalized character on the complex plane as introduced above since $\mathrm{U}(1)_{\mathbb{C}}=\mathbb{C}^{\times}$ which is non-compact similarly as the complexification $\mathrm{SL}(2,\mathbb{C})$ of $\mathrm{SU}(2)$ is non-compact.
\end{remark}
The whole set of generalized characters on $\mathbb{C}$ is probably too large. In fact, according to (\ref{eq:4.1.6}), since the fiducial length $\mathring{\mu}\ell_0$ is a real number, it may be already sufficient to restrict to generalized characters (\ref{eq:4.1.12}) labeled by real numbers $C\in\mathbb{R}$. This corresponds to the requirement of a purely imaginary $\holpar\in i\mathbb{R}$. As we will see in the next section, this choice is consistent with the reality conditions. Hence, following \cite{Wilson-Ewing:2015lia}, we will set $\holpar=i$.\\  
Based on the above observations, in order to construct the algebra corresponding to the purely bosonic degrees of freedom, we define a subalgebra $H_{\mathrm{AP}}(\mathbb{C})\subset H(\mathbb{C})$ of the algebra of holomorphic functions on $\mathbb{C}$, called \emph{almost periodic holomorphic functions}, generated by complex linear combinations of generalized characters of $\mathbb{C}$ labeled by real numbers. That is, a general element $T\in H_{\mathrm{AP}}(\mathbb{C})$ will be of the form
\begin{equation}
T(z)=\sum_{i=1}^N{a_ie^{\mu_i z}}
\end{equation}
with $a_i\in\mathbb{C}$ and $\mu_{i}\in\mathbb{R}$. This can be completed to a Fréchet algebra. Therefore, consider the compact exhaustion $\mathbb{C}=\bigcup_n K_n$ of the complex plane by the compact sets $K_n:=\{z\in\mathbb{C}|\,|z|\leq n\}$, $n\in\mathbb{N}$. The space $H(\mathbb{C})$ of holomorphic functions on $\mathbb{C}$ can then be given the structure of a locally convex space endowing it with the topology of uniform convergence on the $K_n$. In fact, in this way, it turns out that $H(\mathbb{C})$ even has the structure of a \emph{uniform Fréchet algebra} (see \cite{Goldman} and references therein) via pointwise multiplication of holomorphic functions. As a consequence, the closure $\overline{H}_{\mathrm{AP}}(\mathbb{C}):=\overline{H_{\mathrm{AP}}(\mathbb{C})}$ in $H(\mathbb{C})$ inherits the structure of a uniform Fréchet algebra. However, note that it does not define a $*$-algebra. Regarding the standard construction of the state space of LQG (or LQC) via cylindrical functions, we would like to interpret $H_{\mathrm{AP}}(\mathbb{C})$ in terms of (continuous) functions on a group. Therefore, as in case of Banach algebras, one can define the \emph{spectrum} $\mathrm{Spec}\,\overline{H}_{\mathrm{AP}}(\mathbb{C})$ given by the set of all nonzero continuous algebra homomorphisms $\phi:\,\overline{H}_{\mathrm{AP}}(\mathbb{C})\rightarrow\mathbb{C}$. Any $f\in \overline{H}_{\mathrm{AP}}(\mathbb{C})$ canonically induces a linear map on the spectrum via
\begin{equation}
\hat{f}(\phi):=\phi(f),\quad\forall \phi\in\mathrm{Spec}\,\overline{H}_{\mathrm{AP}}(\mathbb{C})
\label{eq:4.1.15}
\end{equation}
called the \emph{Gelfand transform} of $f$. We equip $\mathrm{Spec}\,\overline{H}_{\mathrm{AP}}(\mathbb{C})$ with the \emph{Gelfand topology} given by the coarsest topology such that the Gelfand transforms (\ref{eq:4.1.15}) are continuous. Since $\overline{H}_{\mathrm{AP}}(\mathbb{C})$ is a uniform Fréchet algebra, it follows that $\mathrm{Spec}\,\overline{H}_{\mathrm{AP}}(\mathbb{C})$ is a hemicompact space \cite{Goldman}. As a consequence, the space $C(\mathrm{Spec}\,\overline{H}_{\mathrm{AP}}(\mathbb{C}))$ of continuous functions on the spectrum endowed with the compact open topology also has the structure of a uniform Fréchet algebra. Consider the map
\begin{align}
\Gamma:\,\overline{H}_{\mathrm{AP}}(\mathbb{C})&\rightarrow C(\mathrm{Spec}\,\overline{H}_{\mathrm{AP}}(\mathbb{C}))\\
f&\mapsto\hat{f}\nonumber
\label{eq:eq:4.1.16}
\end{align}    
called the \emph{Gelfand transformation}. It is immediate that $\Gamma$ defines a homomorphism of algebras. In particular, as $\overline{H}_{\mathrm{AP}}(\mathbb{C})$ defines a uniform Fréchet algebra, it follows that $\Gamma$ defines an injective toplogical algebra homomorphism identifying $\overline{H}_{\mathrm{AP}}(\mathbb{C})$ with a closed subalgebra $\Gamma(\overline{H}_{\mathrm{AP}}(\mathbb{C}))$ of $C(\mathrm{Spec}\,\overline{H}_{\mathrm{AP}}(\mathbb{C}))$.\\
The spectrum naturally carries the structure of an abstract group via pointwise multiplication
\begin{align}
\mathrm{Spec}\,\overline{H}_{\mathrm{AP}}(\mathbb{C}))\times\mathrm{Spec}\,\overline{H}_{\mathrm{AP}}(\mathbb{C})&\rightarrow\mathrm{Spec}\,\overline{H}_{\mathrm{AP}}(\mathbb{C})\\
(\phi,\psi)&\mapsto \phi\cdot\psi\nonumber
\label{eq:4.1.17}
\end{align}
and unit $1:\,\overline{H}_{\mathrm{AP}}(\mathbb{C})\rightarrow\mathbb{C},\,f\mapsto 1$. Hence, in this way, we can identify $\overline{H}_{\mathrm{AP}}(\mathbb{C})$ in terms of functions on a group. This is similar to ordinary LQC with real variables. However, it is not clear whether this also forms a topological group. This would follow immediately if, in the definition of $H_{\mathrm{AP}}(\mathbb{C})$, we would have restricted to the subset of ordinary characters $\chi:\,(\mathbb{C},+)\rightarrow\mathrm{U}(1)$. In this case, it follows that the spectrum of the closure has the structure of a (compact, as $\mathrm{U}(1)$ compact) topological group, the so-called Bohr-compactification of the complex plane. We leave it as a task for future investigations to show whether this is also true for $\mathrm{Spec}\,\overline{H}_{\mathrm{AP}}(\mathbb{C})$.\\
\\
With these observations, let us go back to the generalized holonomies $h[\tilde{\mathcal{A}}^+]$ of the super connection. Mimicking the standard procedure of LQC, due to (\ref{eq:4.1.8}) as well as (\ref{eq:4.1.7}), we may identify the matrix elements of the generalized holonomies as functions in
\begin{equation}
H_{\mathrm{AP}}(\mathbb{C})\cup C_0(\mathbb{C})\otimes\Exterior[\phi^A]
\label{eq:4.1.18}
\end{equation}
with $C_0(\mathbb{C})$ the space of continuous functions on $\mathbb{C}$ that vanish at infinity. Interestingly, this is very similar to LQC with real variables. In fact, in \cite{Fleischhack:2015nda} it has been found that these type of functions, already in the pure bosonic sector, i.e. $\phi^A=0$, arise if one considers (bosonic) holonomies along more general edges which are not simply straight edges along integral flows of the left-invariant vector fields, but which may also contain small kinks. Here, we observe that these type of functions appear in the fermionic components of the super holonomies computed along straight edges. \\
\\
In what follows, we will consider a simplified model in which, based on the considerations of \cite{Engle:2016zac}, we will drop functions which are contained in $C_0(\mathbb{C})$. Of course, this means a drastic simplification of the present model which will also break the manifest supersymmetry of the theory. However, this is what is normally done in the literature about canonical minisuperspace models in the framework of supergravity and, as we will see, this model already has a lot of interesting physical implications. We may come back to the more general model for future investigations. Hence, for the rest of this article, we will take as the classical algebra of the theory the following super vector space
\begin{equation}
\mathfrak{A}:=H_{\mathrm{AP}}(\mathbb{C})\otimes\Exterior[\phi^A]
\label{eq:4.1.19}
\end{equation} 
which, in particular, has the structure of a suprcommutative Lie super algebra. To this superalgebra, we need to add the algebra generated by the canonically conjugate momenta both encoded in the super electric field $\mathcal{E}=(p,\pi_{\phi}^{A'})$ (cf. remark \ref{remark:2.1}) which, following the standard procedure in LQG (and LQC), should be implemented as a (super) electric flux operator, i.e., in terms of derivations on $\mathfrak{A}$. More precisely, for any $f\in\mathfrak{A}$, we define
\begin{equation}
p(f)\equiv\{p,f\},\quad\pi_{\phi}^{A'}(f)\equiv\{\pi_{\phi}^{A'},f\}
\label{eq:4.1.20}
\end{equation} 
As the Poisson bracket is a super derivation, it follows that these define elements of the superalgebra $\mathrm{Der}(\mathfrak{A})$ of super derivations on $\mathfrak{A}$. Hence, as the total algebra of the classical theory, we take
\begin{equation}
\widehat{\mathfrak{A}}:=\mathfrak{A}\oplus\mathrm{Der}(\mathfrak{A})
\label{eq:4.1.21}
\end{equation}
This again has the structure of a Lie superalgebra. In fact, $\widehat{\mathfrak{A}}$ defines an ideal w.r.t the action of the super electric fluxes as defined via (\ref{eq:4.1.20}) which, due to the fact the Poisson defines a super right derivations, in particular, defines a representation of Lie superalgebras. Hence, we impose a graded Lie bracket on $\widehat{\mathfrak{A}}$ setting
\begin{equation}
[(f,X),(g,Y)]:=(X(g)-(-1)^{|f||Y|}Y(f),[X,Y])
\end{equation}
for any $(f,X),(g,Y)\in\widehat{\mathfrak{A}}$. So far, $\widehat{\mathfrak{A}}$ does not define a $*$-algebra as the obvious choice of an involution via complex conjugation would lead to anti holomorphic functions and derivations which, for physical reasons, have not been included into the definition of $\widehat{\mathfrak{A}}$. However, one needs to define an involution in order to classify physical quantities in terms of self-adjoint elements. For this reason, let us go back to the reality conditions (\ref{eq:3.3.7.1}) and (\ref{eq:3.3.7.2}). Re-expressing them in terms of the fundamental variables, we may thus impose a $*$-relation on $\widehat{\mathfrak{A}}$ setting
\begin{align}
(e^{\mu\tilde{c}})^{\star}&:=e^{-\mu\tilde{c}}e^{\mu k\ell_0},\quad\quad p^{\star}:=p\\
\phi^{\star}_A&:=i\pi_{\phi}^{A'}n_{AA'},\quad \pi_{\phi}^{\star A}:=i\phi_{A'}n^{AA'}
\end{align}
It is immediate to see that this in fact provides an involution on $\widehat{\mathfrak{A}}$. For instance, one has 
\begin{equation}
(e^{\mu\tilde{c}})^{\star\star}=(e^{-\mu\tilde{c}})^{\star}e^{\mu k\ell_0}=e^{\mu\tilde{c}}e^{-\mu k\ell_0}e^{\mu k\ell_0}=e^{\mu\tilde{c}}
\end{equation}
On the other hand, it follows
\begin{equation}
\phi^{\star\star}_{A'}=-i\pi^{\star A}_{\phi}n_{AA'}=\phi_{B'}n^{AB'}n_{AA'}=\phi_{A'}
\end{equation}
as required. Hence, we have constructed a consistent classcial Lie superalgebra $\widehat{\mathfrak{A}}$ with involution corresponding to chiral $\mathcal{N}=1$ supergravity  which one can interpret as a \emph{graded} analog of the classical \emph{holonomy-flux algebra}.   
\subsection{The kinematical Hilbert space}
\label{sec:rep}
In order to construct the state space of LQSG, we need to find a representation of the graded holonomy-flux algebra $\widehat{\mathfrak{A}}$ on a Hilbert space. More precisely, we are looking for a faithful Lie superalgebra morphism $\pi_{0}$ from $\widehat{\mathfrak{A}}$ to a subset of possibly unbounded operators on a common dense domain of a graded Hilbert space $\mathcal{H}=\mathcal{H}_0\oplus\mathcal{H}_1$ such that $\pi_0(X^{\star})=\pi_0(X)^{\dagger}$ and $\pi_0(f^{\star})=\pi_0(f)^{\dagger}$ $\forall(f,X)\in\widehat{\mathfrak{A}}$. 
\begin{remark}\label{Remark:4.2}
Requiring $\pi_0$ to be faithful, in particular, means that $\pi_0$ preserves the grading. This immediately implies that bosons and fermions \emph{automatically satisfy the correct statistics}. This is in fact a direct consequence of the graded structure of $\widehat{\mathfrak{A}}$ and is rooted in the underlying supersymmetry of the theory since, classically, aupersymmetry requires commuting bosonic and anticommuating fermionic fields.
\end{remark}
As an obvious candidate for a graded pre-Hilbert space $V$, we take 
\begin{equation}
V:=H_{\mathrm{AP}}(\mathbb{C})\otimes\Exterior[\phi^A]
\label{eq:4.2.1}
\end{equation}
Following the standard procedure in LQG and LQC, we define a representation of $\widehat{\mathfrak{A}}$ on this super vector space via
\begin{align}
\widehat{f}:=\pi_0(f)=f,\quad\widehat{p}&:=\pi_0(p):=i\hbar\{p,\cdot\}=-\frac{\hbar\kappa}{3}\frac{\mathrm{d}}{\mathrm{d}\tilde{c}}\label{eq:4.2.2}\\
\widehat{\pi}_{\phi}^{A'}&:=\pi_0(\pi_{\phi}^{A'})=i\hbar\{\pi_{\phi}^{A'},\cdot\}=\frac{\hbar}{i}\frac{\partial}{\partial\phi_{A'}}
\label{eq:4.2.3}
\end{align}
such that the operators corresponding to $f\in\mathfrak{A}$ and the super electric fluxes $p$ and $\pi_{\phi}^{A'}$ act as multiplication operators and derivations, respectively. Requiring commutativity with the involution then implies 
\begin{align}
\widehat{e^{\mu\tilde{c}}}^{\dagger}&=\widehat{e^{-\mu\tilde{c}}}e^{\mu k\ell_0},\quad\quad\widehat{p}^{\dagger}=\widehat{p}\label{eq:4.2.4}\\
\widehat{\phi}^{\dagger}_A&=i\widehat{\pi}_{\phi}^{A'}n_{AA'},\quad\widehat{\pi}_{\phi}^{\dagger A}=i\widehat{\phi}_{A'}n^{AA'}
\label{eq:4.2.5}
\end{align}
$\forall\mu\in\mathbb{R}$. This requires a suitable choice of an inner product on $V$. Hence, we see that the proposal of \cite{Ashtekar:1991hf,Wilson-Ewing:2015lia} to implement the classical reality conditions in terms of a suitable inner product follows very naturally if we interpret them as defining equations for a consistent involution on the classical algebra.\\
To find such an inner product, let us go back to the full theory. As discussed in detail in \cite{Eder:2020erq}, since the theory has underlying $\mathrm{OSp}(1|2)_{\mathbb{C}}$ gauge symmetry, a natural choice of an inner product is given via an invariant measure on $\mathrm{OSp}(1|2)_{\mathbb{C}}$ as a super Lie group. The induced inner product then generically consists of ordinary invariant integral on the bosonic sub Lie group, i.e., $\mathrm{SL}(2,\mathbb{C})$, as well as some variant of a Berezin integral w.r.t. the odd degrees of freedom. Hence, a natural choice for an inner product $\mathscr{S}$ on $V$ is given by
\begin{equation}
\mathscr{S}(f,g):=\int{\mathrm{d}\nu(\tilde{c},\overline{\tilde{c}})\int_B{\mathrm{d}\phi_A\mathrm{d}\bar{\phi}_{A'}\,\bar{f}(\bar{\tilde{c}},\bar{\phi}_A)g(\tilde{c},\phi_{A'})}}
\label{eq:4.2.6}
\end{equation}
where $\int_B$ denotes the standard translation-invariant Berezin integral on $\mathbb{C}^{0|2}$ regarded as purely odd super vector space. This defines a super scalar product on $V$ which, a priori, is indefinite turning $(V,\mathscr{S})$ into an indefinite inner product space. However, it turns out that this can be completed to a Hilbert space. More precisely, as shown in \cite{Tuynman:2018}, one can always find an endomorphism $\mathbf{S}:\,V\rightarrow V$ such that $\mathscr{S}(\cdot,\mathbf{S}\,\cdot)$ defines a positive definite scalar product on $V$. That is, $(V,\mathscr{S})$ is a \emph{Krein space}. This is a standard fact about invariant measures on super Lie groups.\\
The choice of such an endomorphism is are priori completely arbitrary but may be fixed by the requirement of a consistent implementation of the reality conditions (\ref{eq:4.2.5}). A typical choice of $\mathbf{S}$ would be
\begin{equation}
\mathbf{S}:=\exp\left(\frac{1}{\hbar}\bar{\phi}_{A}\phi_{A'}n^{AA'}\right)
\label{eq:4.2.7}
\end{equation}
If $\braket{\cdot\,|\,\cdot}:=\mathscr{S}(\cdot,\mathbf{S}\,\cdot)$, it follows that via the identification $V\cong (H_{\mathrm{AP}}(\mathbb{C}))^{\otimes 4}$, one has
\begin{equation}
\braket{f|g}=\frac{1}{\hbar^2}\braket{\braket{f^0|g^0}}+\frac{1}{\hbar}\braket{\braket{f^+|g^+}}+\frac{1}{\hbar}\braket{\braket{f^-|g^-}}+\braket{\braket{f^2|g^2}}
\label{eq:4.2.8}
\end{equation}
where, for $f\in V$, we made the decomposition $f=f_0+f^{A'}\phi_{A'}+\frac{1}{2}f_2\phi^{A'}\phi_{A'}$ and $\braket{\braket{\cdot\,|\,\cdot}}$ denotes the inner product on $H_{\mathrm{AP}}(\mathbb{C})$, that is,
\begin{equation}
\braket{\braket{f^I|g^I}}:=\int{\mathrm{d}\nu(c,\bar{c})\,\bar{f}^I(\bar{c})g^I(c)}
\label{eq:4.2.9}
\end{equation}
for $I\in\{0,\pm,2\}$. As shown in \cite{DEath:1984gmo,DEath:1996ejf}, with respect to this scalar product, one indeed has
\begin{equation}
\widehat{\phi}^{\dagger}_A=\hbar n_{AA'}\frac{\partial}{\partial\theta_{A'}}=i\widehat{\pi}_{\phi}^{A'}n_{AA'}
\label{eq:4.2.10}
\end{equation}
and thus the reality condition (\ref{eq:4.2.5}) is satisfied. In fact, it turns out that $\mathbf{S}$ is even uniquely fixed by this requirement.\\
Finally, we need to implement the reality conditions (\ref{eq:4.2.4}) for the bosonic degrees of freedom. Since the torsion contribution to the reduced connection simply drops off, this can be done in complete analogy to the matter-free case. Hence, following \cite{Wilson-Ewing:2015lia}, we make the following ansatz
\begin{equation}
\braket{\braket{f|g}}=\lim_{D\rightarrow\infty}\frac{1}{2D}\int_{-D}^D{\mathrm{d}\tilde{c}_R\int_{-D}^D{\mathrm{d}\tilde{c}_I\,\delta\big{(}\tilde{c}_R-\frac{k\ell_0}{2}\big{)}\bar{f}(\bar{\tilde{c}})g(\tilde{c})}}
\label{eq:4.2.11}
\end{equation}
with $f,g\in H_{\mathrm{AP}}(\mathbb{C})$ where we made the decomposition $\tilde{c}=\tilde{c}_R+i\tilde{c}_I$ with $\tilde{c}_R $ and $\tilde{c}_I$ the real and imaginary part of $\tilde{c}$, respectively. Moreover, we have added the spatial curvature $k=0,+1$ of the FLRW spacetime in order to treat both cases simultaneously. Note that, in contrast to \cite{Wilson-Ewing:2015lia}, we do not have to consider the states with positive and negative frequency separately. This is due our sign convention made for the internal 3-form $\epsilon^{ijk}$. Of course, the prize to pay is that then $\tilde{c}$ does not have a straightforward behavior under parity transformations in the purely bosonic case. We however made this choice due to the inclusion of fermionic matter degrees of freedom into the theory.\\
Let us verify that this indeed correctly implements the reality conditions (\ref{eq:4.2.4}). Therefore, note that for elementary states of the form $f=e^{\nu \tilde{c}}$ and $g=e^{\nu' \tilde{c}}$ one has
\begin{align}
\braket{\braket{f|g}}&=\lim_{D\rightarrow\infty}\frac{1}{2D}\int_{-D}^D{\mathrm{d}\tilde{c}_R\int_{-D}^D{\mathrm{d}\tilde{c}_I\,\delta\big{(}\tilde{c}_R-\frac{k\ell_0}{2}\big{)}e^{(\nu+\nu')\tilde{c}_R}e^{i(\nu'-\nu)\tilde{c}_I}}}\nonumber\\
&=e^{(\nu+\nu')\frac{k\ell_0}{2}}\lim_{D\rightarrow\infty}\frac{1}{2D}\int_{-D}^D{\mathrm{d}\tilde{c}_I\,e^{i(\nu'-\nu)\tilde{c}_I}}=e^{\nu k\ell_0}\delta_{\nu,\nu'}
\label{eq:4.2.12}
\end{align}
so that
\begin{equation}
\braket{\braket{f|e^{\mu k\ell_0}e^{-\mu \tilde{c}}g}}=e^{\mu k\ell_0}\braket{\braket{e^{\nu \tilde{c}}|e^{(\nu'-\mu)\tilde{c}}}}=e^{\nu'k\ell_0}\delta_{\mu+\nu,\nu'}=\braket{\braket{e^{\mu\tilde{c}}f|g}}
\label{eq:4.2.13}
\end{equation}
$\forall\mu\in\mathbb{R}$. By linearity, it follows that this holds on all of $V$ so that this indeed provides an implementation of (\ref{eq:4.2.4}). As argued in \cite{Wilson-Ewing:2015lia}, this choice is in fact unique. Thus, we have constructed a unique positive definite inner product on $V$ which we can use to complete $V$ to a Hilbert space $\mathcal{H}'_{\mathrm{kin}}:=\overline{V}^{\|\cdot\|}$. This Hilbert space has the tensor product structure
\begin{equation}
\mathcal{H}'_{\mathrm{kin}}=\mathcal{H}_{\mathrm{grav}}\otimes\mathcal{H}_{f}=\mathcal{H}_{\mathrm{grav}}\otimes\Exterior[\phi_{A'}]
\label{eq:4.2.14}
\end{equation}
and therefore naturally caries a $\mathbb{Z}_2$-grading where $\mathcal{H}_{\mathrm{grav}}$ is the Hilbert space of the purely gravitational degrees of freedom obtained via completion of $H_{\mathrm{AP}}(\mathbb{C})$ w.r.t. the inner product $\braket{\braket{\cdot\,,\,\cdot}}$.\\
According to (\ref{eq:4.2.12}), an orthonormal basis of $\mathcal{H}_{\mathrm{grav}}$ is given by the states $\psi_{\mu}\in\mathcal{H}_{\mathrm{grav}}$ of the form
\begin{equation}
\psi_{\mu}(\tilde{c}):=e^{-\frac{\mu k}{2\ell_0}}e^{\mu \tilde{c}}
\label{eq:4.2.15}
\end{equation}
Since
\begin{equation}
\widehat{p}\psi_{\mu}=-\frac{\hbar\kappa\mu}{3}\psi_{\mu}
\label{eq:4.2.16}
\end{equation}
it follows that these are eigenstates of the bosonic electric flux operator $\widehat{p}$ with eigenvalue $-\frac{\hbar\kappa\mu}{3}\in\mathbb{R}$. Since $\widehat{p}$ factorizes as $\widehat{p}\equiv\widehat{p}\otimes\mathds{1}$ w.r.t. the tensor product structure (\ref{eq:4.2.14}) of the super Hilbert space $\mathcal{H}'_{\mathrm{kin}}$, this implies that $\widehat{p}$ is a densely defined unbounded and symmetric operator on $\mathcal{H}'_{\mathrm{kin}}$ with spectrum contained in the reals, that is, $\widehat{p}$ is self-adjoint. Hence, we have successfully implemented all the reality conditions (\ref{eq:4.2.4}) and (\ref{eq:4.2.5}) on the Hilbert space $\mathcal{H}'_{\mathrm{kin}}$.
\subsection{Solution of the residual Gauss constraint}
We finally have to implement the remaining kinematical constraint given by the residual Gauss constraint (\ref{eq:3.2.12}) given by
\begin{equation}
G_i=\pi_{\phi}^{A'}\tensor{(\tau_i)}{_{A'}^{B'}}\phi_{A'}
\label{eq:4.2.2.1}
\end{equation} 
imposing invariance of the fermionic degrees of freedom under local $\mathrm{SL}(2,\mathbb{C})$ gauge transformations. Note that, due to the hybrid ansatz, it solely depends on the fermionic degrees of freedom as the bosonic variables have been chosen to be isotropic. For the quantization of this constraint, we order $\pi_{\phi}$ to the right so that this yields  
\begin{equation}
\widehat{G}_i=\tensor{(\tau_i)}{_{A'}^{B'}}\widehat{\phi}_{B'}\widehat{\pi}_{\phi}^{A'}
\label{eq:4.2.2.2}
\end{equation} 
We then require that kinemtical states have to be annihiliated by $\widehat{G}_i$. Writing $f=f_0+f^{A'}\phi_{A'}+\frac{1}{2}f_2\phi^{A'}\phi_{A'}$ for a general state $f\in\mathcal{H}_{\mathrm{kin}}$, it follows immediately from the definition that purely bosonic states $f=f_0\in\mathcal{H}'_{\mathrm{kin}}$ are solutions of the constraint equation. For a fully occupied state $f=f_2\phi\phi\in\mathcal{H}'_{\mathrm{kin}}$, it follows
\begin{align}
\widehat{G}_i f_2\phi\phi&=-2i\hbar f_2\tensor{(\tau_i)}{_{A'}^{B'}}\phi_{B'}\phi_{D'}\epsilon^{A'D'}=2i\hbar f_2\tau_i^{B'D'}\phi_{B'}\phi_{D'}\nonumber\\
&=2i\hbar f_2\tau_i^{(B'D')}\phi_{B'}\phi_{D'}=0
\label{eq:4.2.2.3}
\end{align} 
as the fermions are anticommutative while $\epsilon\tau$ is symmetric. On the other hand, for a single fermionic state $f=f^{A'}\phi_{A'}$, one immediately find for $i=3$ that $\widehat{G}_3 f^{A'}\phi_{A'}=0$ if and only if $f^{A'}=0$ $\forall A'$. Hence, solutions to the Gauss constraint are given by kinematical states $f\in\mathcal{H}'_{\mathrm{kin}}$ of the form
\begin{equation}
f=f_0+\frac{1}{2}f_2\phi^{A'}\phi_{A'}
\label{eq:4.2.2.4}
\end{equation}
which is in fact in complete analogy with the results obtained via the approach of D'Eath et al. in \cite{DEath:1996ejf,DEath:1992wve}. States of the form (\ref{eq:4.2.2.4}) form the kinematical Hilbert space which we denote by $\mathcal{H}_{\mathrm{kin}}$. This completes the construction of the kinematical Hilbert space of chiral loop quantum cosmology with local $\mathcal{N}=1$ supersymmetry.
\subsection{The SUSY constraints and the quantum algebra}\label{SUSY constraint}
Having constructed the kinematical Hilbert space of the theory and successfully implemented all the reality conditions, we finally need to determine the physical states. Therefore, note that, according to the constraint algebra (\ref{eq:3.3.21}), the SUSY constraints generate the Hamiltonian constraint. This is a particular property of the canonical description of fields theories with local supersymmetry. This means that, if the SUSY constraints have been succesfully implemented in the quantum theory such that (\ref{eq:3.3.21}) holds in the quantum theory, then these are superior to the Hamiltonian constraint in the sense that, once they are solved, this automatically leads to the solution of the latter. More precisely, if $\Psi\in\mathcal{H}_{\mathrm{kin}}$ with $\widehat{S}^{L A'}\Psi=0=\widehat{S}^{R}_{B'}\Psi$ $\forall A',B'$, then
\begin{equation}
0=[\widehat{S}^{L A'},\widehat{S}^{R}_{B'}]\Psi=\widehat{H}\Psi
\label{eq:4.4.1}
\end{equation}
This is an important feature in canonical quantum supergravity and the Poisson relation (\ref{eq:3.3.21}) poses strong reglementations on the quantization of the constraints and therefore may fix also some of the quantization ambiguities. In \cite{Konsti-SUGRA:2020}, the SUSY constraint has been investigated in the full theory of LQG with real variables. However, the precise relation to the Hamiltonian constraint via studying the quantum algebra has not been considered yet, due to the complexity of the resulting quantum operators. This changes in the symmetry reduced setting where we will be able to study the quantum algebra in detail.\\
In order to implement the SUSY constraints on $\mathcal{H}_{\mathrm{kin}}$, we have to regularize the connection $\tilde{c}$ in the classical expressions, as it not represented as a well-defined operator in the quantum theory. Following \cite{Wilson-Ewing:2015lia}, as typically done in LQC, one therefore fixes some minimal length $\ell_m$ which is of the order of the Planck length and which arises from quantum geometry in the full theory. The connection $\tilde{c}$ will then be approximated via generalized holonomies $e^{\mu\tilde{c}}$ computed along paths of that minimal length $\ell_m$. According to (\ref{eq:4.1.6}), the proper length $\tau$ as measured by the generalized holonomy $e^{\mu\tilde{c}}$ w.r.t. the fiducial metric $\mathring{q}$ is given by $\tau=\mathring{\mu}\ell_0=2\mu\ell_0$. The physical length as measured w.r.t. the metric $a^2\mathring{q}$ is then given by $\tau|a|=2\mu\sqrt{|p|}$ which we require to be the minimal length $\ell_m$ from quantum geometry. Hence, this yields
\begin{equation}
\mu=\frac{\lambda_m}{\sqrt{|p|}}
\label{eq:4.4.2}
\end{equation}
where we set $\lambda_m:=\ell_m/2$. A regularization of the connection $\tilde{c}$ in terms of these generalized holonomies is then given by
\begin{equation}
\tilde{c}=\frac{\sqrt{|p|}}{2\lambda_m}\left(e^{\lambda_m\tilde{c}/\sqrt{|p|}}-e^{-\lambda_m\tilde{c}/\sqrt{|p|}}\right)
\label{eq:4.4.3}
\end{equation}
To study the action of the corresponding operator in the quantum theory, we introduce the new classical variables
\begin{equation}
\beta:=\frac{\widetilde{c}}{\sqrt{|p|}}\quad\text{and}\quad V:=\epsilon|p|^{\frac{3}{2}}
\label{eq:4.4.4}
\end{equation}
which satisfy the classical Poisson bracket
\begin{equation}
\{\beta,V\}=-\frac{i\kappa}{2}
\label{eq:4.4.5}
\end{equation}
and thus defines a canonically conjugate pair. From this, it follows
\begin{equation}
\{e^{\lambda_m\beta},V\}=\sum_{k=1}^{\infty}\frac{(\lambda_m)^k}{k!}\{\beta^k,V\}=-\frac{i\kappa\lambda_m}{2}e^{\lambda_m\beta}
\label{eq:4.4.6}
\end{equation}
On $\mathcal{H}_{\mathrm{kin}}$, let us introduce the states
\begin{equation}
\ket{V}:=e^{\frac{k\mu(V)}{2\ell_0}}\psi_{\mu(V)},\quad\text{with}\quad\mu(V):=-\frac{3}{\hbar\kappa}\mathrm{sign}(V)|V|^{\frac{2}{3}}
\label{eq:4.4.7}
\end{equation}
for $V\in\mathbb{R}$ where $\psi_{\mu}$ is the orthonormal basis of $\mathcal{H}_{\mathrm{grav}}$ as defined via (\ref{eq:4.2.15}). The states (\ref{eq:4.4.7}) are eigenstates of the volume operator $\widehat{V}:=\epsilon|\widehat{p}|^{\frac{3}{2}}$ with eigenvalue $V$. Note that they are normalized only case of vanishing spatial curvature. We will however use them for both cases as the action of the holonomy operators on these states is much simpler \cite{Wilson-Ewing:2015lia}. According to (\ref{eq:4.4.6}), it follows\footnote{as common in the LQC literature, for notational simiplification, we will drop hats indicating (bosonic) operator expressions in what follows}
\begin{equation}
V\left(e^{\lambda_m\beta}\ket{V}\right)=[V,e^{\lambda_m\beta}]\ket{V}+V e^{\lambda_m\beta}\ket{V}=\left(V-\frac{\hbar\kappa\lambda_m}{2}\right)e^{\lambda_m\beta}\ket{V}
\end{equation}
Setting $v:=\frac{4}{\hbar\kappa}V$, we thus obtain
\begin{equation}
e^{\lambda_m\beta}\ket{v}=\ket{v-2\lambda_m}
\label{eq:4.4.8}
\end{equation}
For the implementation of the SUSY constraints (\ref{eq:3.3.9}) and (\ref{eq:3.3.10}) as well as the Hamiltonian constraint (\ref{eq:3.3.8}), we exploit quantization ambiguities to quantize expressions of the form $\epsilon\tilde{c}$ in the most symmetric way. In fact, as we will see, symmetric ordering will lead to a correct implementation of the classical constraint algebra (\ref{eq:3.3.21}). Hence, using the regularisation (\ref{eq:4.4.3}), for the quantum analog of $\epsilon\tilde{c}$, we follow \cite{MartinBenito:2009aj} and take
\begin{equation}
\epsilon\widetilde{c}=\frac{\sqrt{|p|}}{2\lambda_m}\left(\mathcal{N}_{-}-\mathcal{N}_{+}\right)
\label{eq:4.4.9}
\end{equation}
where
\begin{equation}
\mathcal{N}_{\pm}:=\frac{1}{2}\left(e^{\mp\lambda_m\widetilde{c}/\sqrt{|p|}}\epsilon+\epsilon e^{\mp\lambda_m\widetilde{c}/\sqrt{|p|}}\right)
\label{eq:4.4.10}
\end{equation}
Due to (\ref{eq:4.4.8}), the action of these operators on volume eigenstates are given by
\begin{equation}
\mathcal{N}_{\pm}\ket{v}=\frac{1}{2}\big{(}\mathrm{sign}(v)+\mathrm{sign}(v\pm 2)\big{)}\ket{v\pm 2\lambda_m}
\label{eq:4.4.11}
\end{equation}
and thus $\mathcal{N}_{-}\ket{v}$ resp. $\mathcal{N}_{+}\ket{v}$ vanishes in case $v\in(0,2\lambda_m)$ resp. $v\in(-2\lambda_m,0)$. Next, we have to implement the classical quantity $i\phi_{\phi}^{A'}\phi_{A'}$ which appears, for instance, in equation (\ref{eq:3.3.4}) relating the reduced connections $c$ and $\tilde{c}$  after having performed a canonical transformation to half-densitized fermionic fields. Sticking again to symmetric ordering, we define  
\begin{equation}
\widehat{\Theta}:=\frac{i}{2}(\widehat{\pi}^{A'}_{\phi}\widehat{\phi}_{A'}-\widehat{\phi}_{A'}\widehat{\pi}^{A'}_{\phi})=\hbar-i\widehat{\phi}_{A'}\widehat{\pi}^{A'}_{\phi}
\label{eq:4.4.12}
\end{equation}
By definition, it follows that $\widehat{\Theta}$ is self-adjoint, since, due to reality conditions (\ref{eq:4.2.5}),
\begin{equation}
\left(i\widehat{\phi}_{A'}\widehat{\pi}^{A'}_{\phi}\right)^{\dagger}=-i\widehat{\pi}^{\dagger A}_{\phi}\widehat{\phi}^{\dagger}_{A}=i\widehat{\phi}_{B'}\widehat{\pi}_{\phi}^{C'}n^{AB'}n_{AC'}=i\widehat{\phi}_{A'}\widehat{\pi}^{A'}_{\phi}
\label{eq:4.4.13}
\end{equation}
Moreover, $\widehat{\Theta}$ annihilates states with one single fermionic excitation, that is, $\widehat{\Theta}\phi_{A'}=0$. In fact, $\widehat{\Theta}$ is related to the particle number operator $\widehat{N}$ via $\widehat{N}=\mathds{1}-\frac{1}{\hbar}\widehat{\Theta}$, where $\widehat{N}$ counts the number of fermionic excitations in a quantum state. Using (\ref{eq:4.4.12}) as well as (\ref{eq:4.4.9}), it follows that the quantum analog of (\ref{eq:3.3.4}) takes the form 
\begin{align}
\epsilon c&=\epsilon\widetilde{c}-\frac{\kappa\epsilon}{12 p}\frac{i}{2}(\widehat{\pi}^{A'}_{\phi}\widehat{\phi}_{A'}-\widehat{\phi}_{A'}\widehat{\pi}^{A'}_{\phi})\nonumber\\
&=\frac{g^{\frac{1}{3}}|v|^{\frac{1}{3}}}{2\lambda_m}\left((\mathcal{N}_{-}-\mathcal{N}_{+})-\frac{\kappa\lambda_m}{6g|v|}\widehat{\Theta}\right)
\label{eq:4.4.14}
\end{align}
where we used that $|p|=(2\pi G\hbar)^{\frac{2}{3}}|v|^{\frac{2}{3}}=:g^{\frac{2}{3}}|v|^{\frac{2}{3}}$. To implement the dynamical constraints in the quantum theory, for sake of simplicity, we want to restrict to the special case $k=0$ and $L\rightarrow\infty$, i.e., vanishing spatial curvature and cosmological constant. This will simplify the derivation of the quantum algebra and also indicate very clearly how the classical algebra can be maintained in the quantum theory. The case with non-vanishing spatial curvature will be discussed in the following section in the context of the semi-classical limit of the theory.\\
If we use symmetric ordering, it follows that the Hamiltonian constraint operator in the quantum theory can be defined in the following way
\begin{align}
\widehat{H}=&-\frac{3g}{4\kappa\lambda_m^2}|v|^{\frac{1}{4}}\left((\mathcal{N}_{-}-\mathcal{N}_{+})-\frac{\kappa\lambda_m}{6g|v|}\widehat{\Theta}\right)|v|^{\frac{1}{2}}\left((\mathcal{N}_{-}-\mathcal{N}_{+})-\frac{\kappa\lambda_m}{6g|v|}\widehat{\Theta}\right)|v|^{\frac{1}{4}}\nonumber\\
&-\frac{1}{8\lambda_m}|v|^{\frac{1}{4}}\left((\mathcal{N}_{-}-\mathcal{N}_{+})-\frac{\kappa\lambda_m}{6g|v|}\widehat{\Theta}\right)|v|^{-\frac{1}{4}}\widehat{\Theta}\nonumber\\
&-\frac{1}{8\lambda_m}|v|^{-\frac{1}{2}}|v|^{\frac{1}{4}}i\widehat{\pi}^{A'}_{\phi}\left((\mathcal{N}_{-}-\mathcal{N}_{+})-\frac{\kappa\lambda_m}{6g|v|}\widehat{\Theta}\right)|v|^{\frac{1}{4}}\widehat{\phi}_{A'}
\label{eq:4.4.15}
\end{align}
The bosonic part of $\widehat{H}$ is almost the same as in \cite{Wilson-Ewing:2015lia}. However, unlike in \cite{Wilson-Ewing:2015lia}, we did not change the overall power of the rescaled volume operator by redefining the lapse function $N$. This is due to the fact that the Hamiltonian constraint should be related to the SUSY constraints via a quantum analog of (\ref{eq:3.3.21}). Hence, redefining the Hamiltonian constraint requires a redefinition of the SUSY constraints which may then change the resulting quantum algebra which we would like avoid.\\
Note that each of the two fermionic contributions in (\ref{eq:4.4.15}) indeed yield half of $H_f$ in the classical limit $\hbar\rightarrow 0$ in which case the terms $|v|^{\frac{1}{4}}$ and $|v|^{-\frac{1}{4}}$ simply cancel each other. Moreover, in the last line of (\ref{eq:4.4.15}),  the term $|v|^{-\frac{1}{4}}$ on the left hand side of the bracket has been regularized replacing it by the equivalent expression $|v|^{-\frac{1}{2}}|v|^{\frac{1}{4}}$ for $v\neq 0$. In this way, it follows that the Hamiltonian constraint is well-defined by acting on all states $\ket{v}$ with $v\neq 0$. In fact, suppose $\widehat{H}$ acts on the state $\ket{v}=\ket{2}$, then this state will mapped to zero-volume state $\ket{0}$ by the shift operator $\mathcal{N}_-$ which will subsequently be annihilated by $|v|^{\frac{1}{2}}$. A similar kind of reasoning applies to the sector with $v<0$. Hence, the Hamiltonian has a well-defined action on the states $\ket{v}=\ket{\pm 2}$.\\
As will become clear in what follows, this kind of regularisation of the Hamiltonian constraint operator is not imposed artificially but turns out to be even mandatory if one requires consistency with the classical Poisson relation (\ref{eq:3.3.21}). Therefore, let us implement the SUSY constraints in the quantum theory. If we stick to symmetric ordering, this immediately gives  
\begin{align}
\widehat{S}^{L A'}=\frac{g^{\frac{1}{2}}}{2\lambda_m}|v|^{\frac{1}{4}}\left((\mathcal{N}_{-}-\mathcal{N}_{+})-\frac{\kappa\lambda_m}{6g|v|}\widehat{\Theta}\right)|v|^{\frac{1}{4}}\widehat{\pi}^{A'}_{\phi}
\label{eq:4.4.16}
\end{align}
for the left supersymmetry constraint as well as
\begin{align}
\widehat{S}^{R}_{A'}=&\frac{3g^{\frac{1}{2}}}{2\lambda_m}|v|^{\frac{1}{4}}\left((\mathcal{N}_{-}-\mathcal{N}_{+})-\frac{\kappa\lambda_m}{6g|v|}\widehat{\Theta}\right)|v|^{\frac{1}{4}}\widehat{\phi}_{A'}
\label{eq:4.4.17}
\end{align}
for the right supersymmetry constraint. By definition, these operators are both well-defined while acting on states $\ket{v}$ with $v\neq 0$. For the computation of the quantum algebra among the SUSY constraints, we have to calculate various commutators of the form $[\widehat{\Theta},\widehat{\pi}_{\phi}^{A'}]$ and $[\widehat{\Theta},\widehat{\phi}_{A'}]$ which are given by
\begin{equation}
[\widehat{\Theta},\widehat{\pi}_{\phi}^{A'}]=\hbar\widehat{\pi}_{\phi}^{A'}\quad\text{and}\quad[\widehat{\Theta},\widehat{\phi}_{A'}]=-\hbar\widehat{\phi}_{A'}
\label{eq:4.4.18}
\end{equation} 
Thus, using (\ref{eq:4.4.18}), we find that the trace part of the operator-valued matrix $[\widehat{S}^{L A'},\widehat{S}^{R}_{B'}]$ is given by

\begin{align}
[\widehat{S}^{L A'},\widehat{S}^{R}_{A'}]&=\frac{3g}{4\lambda_m^2}|v|^{\frac{1}{4}}\left((\mathcal{N}_{-}-\mathcal{N}_{+})-\frac{\kappa\lambda_m}{6g|v|}\widehat{\Theta}\right)|v|^{\frac{1}{2}}\left((\mathcal{N}_{-}-\mathcal{N}_{+})-\frac{\kappa\lambda_m}{6g|v|}\widehat{\Theta}\right)|v|^{\frac{1}{4}}[\widehat{\pi}^{A'}_{\phi},\widehat{\phi}_{A'}]\nonumber\\
&+\frac{3g}{4\lambda_m^2}|v|^{\frac{1}{4}}\left((\mathcal{N}_{-}-\mathcal{N}_{+})-\frac{\kappa\lambda_m}{6g|v|}\widehat{\Theta}\right)|v|^{\frac{1}{4}}\left[\widehat{\pi}^{A'}_{\phi},|v|^{\frac{1}{4}}\left((\mathcal{N}_{-}-\mathcal{N}_{+})-\frac{\kappa\lambda_m}{6g|v|}\widehat{\Theta}\right)|v|^{\frac{1}{4}}\right]\widehat{\phi}_{A'}\nonumber\\
&-\frac{3g}{4\lambda_m^2}|v|^{\frac{1}{4}}\left((\mathcal{N}_{-}-\mathcal{N}_{+})-\frac{\kappa\lambda_m}{6g|v|}\widehat{\Theta}\right)|v|^{\frac{1}{4}}\left[|v|^{\frac{1}{4}}\left((\mathcal{N}_{-}-\mathcal{N}_{+})-\frac{\kappa\lambda_m}{6g|v|}\widehat{\Theta}\right)|v|^{\frac{1}{4}},\widehat{\phi}_{A'}\right]\widehat{\pi}^{A'}_{\phi}\nonumber\\
&=2i\hbar\kappa\widehat{H}_b-\frac{\kappa}{8\lambda_m}|v|^{\frac{1}{4}}\left((\mathcal{N}_{-}-\mathcal{N}_{+})-\frac{\kappa\lambda_m}{6g|v|}\widehat{\Theta}\right)|v|^{-\frac{1}{4}}[\widehat{\pi}_{\phi}^{A'},\widehat{\Theta}]\widehat{\phi}_{A'}\nonumber\\
&+\frac{\kappa}{8\lambda_m}|v|^{\frac{1}{4}}\left((\mathcal{N}_{-}-\mathcal{N}_{+})-\frac{\kappa\lambda_m}{6g|v|}\widehat{\Theta}\right)|v|^{-\frac{1}{4}}[\widehat{\Theta},\widehat{\phi}_{A'}]\widehat{\pi}_{\phi}^{A'}\nonumber\\
&=2i\hbar\kappa\left(\widehat{H}_b-\frac{1}{8\lambda_m}|v|^{\frac{1}{4}}\left((\mathcal{N}_{-}-\mathcal{N}_{+})-\frac{\kappa\lambda_m}{6g|v|}\widehat{\Theta}\right)|v|^{-\frac{1}{4}}\widehat{\Theta}\right)\nonumber\\
&=2i\hbar\kappa\widehat{H}-\frac{\hbar\kappa}{4\lambda_m|v|^{\frac{1}{2}}}\widehat{\pi}_{\phi}^{A'}\left(|v|^{\frac{1}{4}}\left((\mathcal{N}_{-}-\mathcal{N}_{+})-\frac{\kappa\lambda_m}{6g|v|}\widehat{\Theta}\right)|v|^{\frac{1}{4}}\widehat{\phi}_{A'}\right)
\label{eq:4.4.19}
\end{align}
that is
\begin{align}
[\widehat{S}^{L A'},\widehat{S}^{R}_{A'}]=2i\hbar\kappa\widehat{H}-\frac{\hbar\kappa}{6g^{\frac{1}{2}}|v|^{\frac{1}{2}}}\widehat{\pi}_{\phi}^{A'}\widehat{S}^{R}_{A'}
\label{eq:4.4.20}
\end{align}
which is precisely the quantum analog of (\ref{eq:3.3.16}). To compute the off-diagonal entries of $[\widehat{S}^{L A'},\widehat{S}^{R}_{B'}]$ note that, by definition, the right SUSY constraint operator $\widehat{S}_{B'}^R$ creates a fermionic state $\phi_{B'}$. On the other hand, $\widehat{S}^{L A'}$ annihilates a fermionic state labelled with $A'$. Hence, if $[\widehat{S}^{L A'},\widehat{S}^{R}_{B'}]$ acts on a fermionic vacuum state $f=f_0\in\mathcal{H}_{\mathrm{kin}}$, this immediately gives for $A'\neq B'$ 
\begin{equation}
[\widehat{S}^{L A'},\widehat{S}^{R}_{B'}]f_0=\widehat{S}^{L A'}(\widehat{S}^{R}_{B'}f_0)=0=\frac{\hbar\kappa}{6g^{\frac{1}{2}}|v|^{\frac{1}{2}}}\widehat{\pi}_{\phi}^{A'}(\widehat{S}^{R}_{B'}f_0)
\label{eq:4.4.21}
\end{equation}
On the other hand, in case $f=f_2\phi\phi\in\mathcal{H}_{\mathrm{kin}}$ is a fully occupied state, it follows
 \begin{equation}
[\widehat{S}^{L A'},\widehat{S}^{R}_{B'}]f=-\widehat{S}^{R}_{B'}(\widehat{S}^{L A'}f_2\phi\phi)=0=\frac{\hbar\kappa}{6g^{\frac{1}{2}}|v|^{\frac{1}{2}}}\widehat{\pi}_{\phi}^{A'}(\widehat{S}^{R}_{B'}f)
\label{eq:4.4.22}
\end{equation}
since $\widehat{S}^{R}_{B'}(\widehat{S}^{L A'}f_2\phi\phi)\propto (\phi_{B'})^2=0$ for $A'\neq B'$ by anticommutativity of the fermions. Hence, to summarize, we found that the quantum algebra between the left and right supersymmetry constraint on $\mathcal{H}_{\mathrm{kin}}$ takes the form
\begin{equation}
[\widehat{S}^{L A'},\widehat{S}^{R}_{B'}]=\left(i\hbar\kappa\widehat{H}-\frac{\hbar\kappa}{6g^{\frac{1}{2}}|v|^{\frac{1}{2}}}\widehat{\pi}_{\phi}^{A'}\widehat{S}^{R}_{A'}\right)\delta^{A'}_{B'}+\frac{\hbar\kappa}{6g^{\frac{1}{2}}|v|^{\frac{1}{2}}}\widehat{\pi}_{\phi}^{A'}\widehat{S}^{R}_{B'}
\label{eq:4.4.23}
\end{equation}
and thus exactly reproduces the classical Poisson relation (\ref{eq:3.3.21}). This also justifies the symmetric ordering chosen for the Hamiltonian constraint \cite{MartinBenito:2009aj}. This is in fact a standard quantization scheme used in LQC and follows here requiring consistency with the SUSY constraints.\\
\\
As a final step, one needs to find physical states in $\mathcal{H}_{\mathrm{kin}}$ annihilated by the constraint operators in order to study the dynamics of the theory. As already explained above, it therefore suffices to solve the SUSY constraints as, via (\ref{eq:4.4.23}), this immediately leads to a solution of the Hamiltonian constraint. However, in order to introduce a relational clock, one cannot simply add a scalar field to the theory as usually done in LQC. This is due to the fact, that ordinary scalar fields will not contribute to the SUSY constraints leading to inconsistent dynamics according to (\ref{eq:4.4.23}). Hence, instead, one needs to consider local supersymmetric matter coupled to supergravity, i.e., scalar fields with additional spin-$\frac{1}{2}$ matter fields.
\begin{remark}
\label{rem_classical}
Let us briefly comment on the dynamics of the classical theory and the suitability of fermionic fields as relational clocks (see also remark \ref{rem_clock} below). As mentioned already in remark \ref{Remark:4.2}, it is an immediate consequence of supersymmetry that classical fermionic fields have to be anticommuting which may be regarded as the classical limit of the spin-statistics theorem in quantum field theory. As a result, the fermionic fields are \emph{odd Grassmann-valued} functions on phase space and thus, in particular, are nilpotent. Generally, any Grassmann number has an ordinary real (resp. complex) valued scalar coefficient at lowest order which is typically referred to as its \emph{body}. These are the numbers one has access to in classical experiments. However, as fermionic fields are Grassman-odd, it follows that their body has to be zero. For this reason, fermionic fields do not serve as good 'clocks' in cosmology even at the classical level.\\
Taking the body of the supergravity action \eqref{eq:0.2.1}, the fermionic contributions immediately drop off due to nilpotency leading back to the standard Palatini action of first-order Einstein gravity with cosmological constant. Hence, classically, one may interpret this limit in terms of the evolution of fermions on the classical bosonic background.\\
Nevertheless, one may consider instead bosonic quantities (or rather their expectation values) derived from the fermionic fields. For instance, one can consider the current $J_0:=i\pi^{A'}_{\phi}\phi_{A'}$ or the associated fermionic energy density $\rho_f$ which, in case of a vanishing cosmological constant, is related to the current via\footnote{Probably, the easiest way to see this is to use the Hamiltonian constraint $H\approx 0$ from which one can read off the imaginary part $c_I=\Im(c)$ of the symmetry reduced Ashtekar connection. Since $\dot{a}=\{a,H\}=c_I$ this yields an expression for the \emph{Hubble parameter} $\dot{a}/a$ which, via the \emph{Friedmann equation} for a homogeneous isotropic universe, is related to the fermionic energy density via $(\dot{a}/a)^2=\kappa\rho_f/3-(k\ell_0)^2/4a^2$.} $\rho_f\sim J_0^2/a^6$. Further bosonic quantities, in case of a non-vanishing cosmological constant, are given by the, in general complex, currents $J_{\phi}:=i\epsilon^{A'B'}\phi_{A'}\phi_{B'}$ and $\bar{J}_{\phi}:=-i\epsilon_{A'B'}\pi^{A'}_{\phi}\pi^{B'}_{\phi}$, respectively. The classical Hamiltonian constraint may then be written as
\begin{align}
H=&-\frac{3\epsilon^2}{\kappa }\sqrt{|p|}(c^2-k\ell_0 c)-\frac{\epsilon}{2\sqrt{|p|}}(c-k\ell_0)J_0-\frac{\epsilon^2}{L}\Re(J_{\phi})+\frac{3}{\kappa L^2}|p|^{\frac{3}{2}}\nonumber
\end{align}
However, in the classical theory and without coupling it to additional locally supersymmetric matter fields, it turns out to be hard to find nontrivial solutions of this pure graviton-gravitino model with non-vanishing fermion currents which lead to a non-static dynamical universe (this is mainly due to the constraints imposed by the SUSY and Hamiltonian constraints on the initial conditions). This coincides with the observations in \cite{DEath:1996ejf,DEath:1992wve}, where it is argued that this pure graviton-gravitino model has to be considered quantum theoretically. Hence, for various reasons, it is desirable to study the system coupled to further locally supersymmetric matter fields. According to the discussions in \cite{DEath:1996ejf}, this may in fact lead to very interesting dynamics.    
\end{remark}
\begin{remark}
\label{rem_clock}
As already argued in the previous remark, even in the classical theory, fermionic fields do not serve as good relational clocks. One may then consider instead derived bosonic quantities such as the fermionic current $J_0$ which in the quantum theory, up to ordering ambiguities, is represented by the fermionic particle number operator $\widehat{N}=\mathds{1}-\frac{1}{\hbar}\widehat{\Theta}$. However, this operator has pure discrete spectrum consisting of the three eigenvalues $\{0,1,2\}$. Hence, in the quantum theory, this quantity also does not serve as a good relational clock.
Alternatively, one may use $p$ as a clock. The fermionic state then becomes a function of this ``gravitational'' time. We have done something like this in the discussion of the semi-classical limit in section \ref{sec:semi} below.
\end{remark}
Based on the observations in remark \ref{rem_classical} and \ref{rem_clock}, we leave it as a task for future investigations to study the full dynamics of the theory including local supersymmetric matter fields as a relational clock. Nevertheless, one can already make some qualitative statements concerning the singularity resolution. In fact, according to (\ref{eq:4.4.16}) and (\ref{eq:4.4.17}), by acting with the quantum SUSY constraints, irrespective of the number of fermionic excitations, the volume eigenstates $\ket{v}=\ket{\pm 2}$ are mapped to the zero volume state $\ket{0}$ or $\ket{\pm 4}$ by the shift operators $\mathcal{N}_{\pm}$. The zero volume state is then subsequently annihilated by the volume operator $|v|^{\frac{1}{4}}$. Moreover, states $\ket{v}$ with $v\in(0,2\lambda_m)$ are annihilated by $\mathcal{N}_-$ whereas states $\ket{v}$ with $v\in(-2\lambda_m,0)$ are mapped to zero by $\mathcal{N}_+$. Hence, it follows that the vacuum state decouples from the dynamics and accordingly the cosmic singularity is resolved in this model.
\subsection{The semi-classical limit}\label{sec:semi}
So far, for the derivation of the quantum algebra (\ref{eq:4.4.23}), we have restricted to the special case of a vanishing spatial curvature and vanishing cosmological constant. In order to compare our model with other supersymmetric minisuperspace models in the literature, we next want to consider the case of a positive spatial curvature, i.e., $k=1$, and study the semi-classical limit of the theory. As observed already in the previous section, in order to ensure consistency with classical Poisson algebra, it is worthwhile to choose a symmetric ordering for the dynamical constraints. Hence, we define  
\begin{align}
\widehat{S}^{L A'}=\frac{g^{\frac{1}{2}}}{2\lambda_m}|v|^{\frac{1}{4}}\left((\mathcal{N}_{-}-\mathcal{N}_{+})-\frac{\kappa\lambda_m}{6g|v|}\widehat{\Theta}\right)|v|^{\frac{1}{4}}\widehat{\pi}^{A'}_{\phi}
\label{eq:4.5.1}
\end{align}
for the left supersymmetry constraint which is equivalent to (\ref{eq:4.4.17}) and respectively
\begin{align}
\widehat{S}^{R}_{A'}=&\frac{3g^{\frac{1}{2}}}{2\lambda_m}|v|^{\frac{1}{4}}\left((\mathcal{N}_{-}-\mathcal{N}_{+})-\frac{\kappa\lambda_m}{6g|v|}\widehat{\Theta}\right)|v|^{\frac{1}{4}}\widehat{\phi}_{A'}-3\epsilon g^{\frac{1}{6}}|v|^{\frac{1}{6}}\ell_0\widehat{\phi}_{A'}
\label{eq:4.5.2}
\end{align}
for the right supersymmetry constraint. Since $\widehat{S}^{L A'}$ and $\widehat{S}^{R}_{B'}$ for $A'\neq B'$ create respectively annihilate fermions with opposite quantum numbers, as in the previous section, it is immediate to see that on the subspace of gauge-invariant states the commutator $[\widehat{S}^{L A'},\widehat{S}^{R}_{B'}]$ simply vanishes such that
\begin{equation}
[\widehat{S}^{L A'},\widehat{S}^{R}_{B'}]=\frac{\hbar\kappa}{6g^{\frac{1}{2}}|v|^{\frac{1}{2}}}\widehat{\pi}_{\phi}^{A'}\widehat{S}^{R}_{A'}=0
\label{eq:4.5.3}
\end{equation}
on $\mathcal{H}_{\mathrm{kin}}$ for $A'\neq B'$ as required. The trace part can then be considered as defining equation of the Hamiltonian constraint. We will not derive an explicit expression of Hamiltonian constraint in what follows. Instead, we want to turn to the semi-classical limit of the theory. Therefore, note that the minimal length $\lambda_m$ (resp. $\ell_m$) should arise from quantum geometry of the full theory of self-dual LQG. More precisely, it should be related to the discrete spectrum of the area operator in terms of the square root of the minimal possible area eigenvalue. Hence, following \cite{Wilson-Ewing:2015lia,Rovelli:1994ge}, this suggests that $\lambda_m^2=16\sqrt{3}\pi G\hbar$. \\
In this section, we are interested in the limit in which effects from quantum geometry are negligible, that is, in which case the quantum area spectrum becomes nearly continuous. Hence, we consider the limit $\lambda_m\rightarrow 0$. \\     
Since the SUSY constraints are superior to the Hamiltonian constraint in canonical quantum supergravity, it suffices to find the semi-classical solutions to (\ref{eq:4.5.1}) and (\ref{eq:4.5.2}). Hence, let us consider a state $\Psi\in\mathcal{H}_{\mathrm{kin}}$ of the form   
\begin{equation}
\Psi=\sum_{v}\psi(v)\ket{v}+\left(\sum_{v}\psi'(v)\ket{v}\right)\otimes\phi^{A'}\phi_{A'}
\label{eq:4.5.4}
\end{equation}
with certain complex coefficients $\psi(v)$ and $\psi'(v)$ which we assume to correspond to at least once continuously differentiable functions $\psi:\,v\mapsto\psi(v)$ and $\psi':\,v\mapsto\psi'(v)$ on some open subset of $\mathbb{R}_{>0}$ where we have restricted to the sector of positive volume. It follows that $\Psi$ is a physical state, that is, $\Psi\in\mathcal{H}_{\mathrm{phys}}$, if and only if $\widehat{S}^{L A'}\Psi=0=\widehat{S}^R_{B'}\Psi$ $\forall A',B'$. Concerning the right supersymmetry constraint, this yields, using the fact that $\widehat{\Theta}$ annihilates states with one single fermionic excitation,
\begin{align}
0=\widehat{S}^{R}_{A'}\Psi=&\frac{3g^{\frac{1}{2}}}{2\lambda_m}\sum_v\left(|v|^{\frac{1}{4}}|v-2\lambda_m|^{\frac{1}{4}}\psi(v)\ket{v-2\lambda_m}-|v|^{\frac{1}{4}}|v+2\lambda_m|^{\frac{1}{4}}\psi(v)\ket{v+2\lambda_m}\right.\nonumber\\
&\quad\quad\left.-3g^{\frac{1}{6}}|v|^{\frac{1}{6}}\ell_0\psi(v)\ket{v}\right)\otimes\phi_{A'}\nonumber\\
=&\frac{3g^{\frac{1}{2}}}{2\lambda_m}\sum_v\bigg{[}|v+2\lambda_m|^{\frac{1}{4}}|v|^{\frac{1}{4}}\psi(v+2\lambda_m)-|v-2\lambda_m|^{\frac{1}{4}}|v|^{\frac{1}{4}}\psi(v-2\lambda_m)\nonumber\\
&\quad\quad-3g^{\frac{1}{6}}|v|^{\frac{1}{6}}\ell_0\psi(v)\bigg{]}\ket{v}\otimes\phi_{A'}
\label{eq:4.5.5}
\end{align}
where we have performed an index shift in order to obtain the last line. Hence, setting $F(v):=|v|^{\frac{1}{4}}\psi(v)$, it follows that $\Psi$ is a solution to the right SUSY constraint operator if and only if $F$ satisfies the difference equation
\begin{equation}
\frac{F(v+2\lambda_m)-F(v-2\lambda_m)}{4\lambda_m}=\frac{\ell_0}{2}\frac{1}{g^{\frac{1}{3}}|v|^{\frac{1}{3}}}|v|^{\frac{1}{4}}\psi(v)=\frac{\ell_0}{2}\frac{1}{g^{\frac{1}{3}}|v|^{\frac{1}{3}}}F(v)
\label{eq:4.5.6}
\end{equation}
Thus, in the limit where effects from quantum geometry are negligible, one can approximate the difference on the l.h.s. of equation (\ref{eq:4.5.6}) by a derivative yielding  
\begin{equation}
F'(v)=\lim_{\lambda_m\rightarrow 0}\frac{F(v+2\lambda_m)-F(v-2\lambda_m)}{4\lambda_m}=\frac{\ell_0}{2}\frac{1}{g^{\frac{1}{3}}|v|^{\frac{1}{3}}}F(v)
\label{eq:4.5.7}
\end{equation}
which after separating variables and integrating on both sides gives
\begin{equation}
\mathrm{ln}\,F(v)=\frac{3\ell_0}{4g^{\frac{1}{3}}}|v|^{\frac{2}{3}}=\frac{3a^2V_0}{\kappa\hbar}+C
\label{eq:4.5.8}
\end{equation}
where $C$ is some constant fixed by the initial conditions. Hence, it follows that in the semi-classical limit the unique solution to the right SUSY constraint, up to a constant multiple, is given by 
\begin{equation}
F(v)=\exp\left(\frac{3a^2V_0}{\kappa\hbar}\right)\quad\Leftrightarrow\quad \psi(v)=|v|^{-\frac{1}{4}}\exp\left(\frac{3a^2V_0}{\kappa\hbar}\right)
\label{eq:4.5.9}
\end{equation}
Finally, the solution $\widehat{S}^{L A'}\Psi=0$ to the left SUSY constraint are obtained following the same steps as before which yields $\psi'(v)=|v|^{-\frac{1}{4}}$. Hence, we found a general solution of the quantum SUSY constraints in the semi-classical limit $\lambda_m\rightarrow 0$ in which effects from quantum geometry are negligible take the form
\begin{equation}
C|v|^{-\frac{1}{4}}\exp\left(\frac{3a^2V_0}{\kappa\hbar}\right)+D|v|^{-\frac{1}{4}}\phi^{A'}\phi_{A'}
\label{eq:4.5.10}
\end{equation}  
with some constant coefficients $C,D\in\mathbb{C}$. For $D=0$ the exponential term is exactly the semi-classical state as found in \cite{DEath:1996ejf,DEath:1992wve} and, as will be explained in more detail below, turns out to be a \emph{Hartle-Hawking} state of the theory. In the present case, the additional correction term $|v|^{-\frac{1}{4}}$ arises from the symmetric ordering chosen for the definition of the quantum SUSY constraints and thus is simply a quantization ambiguity. In fact, if the regularized connection in (\ref{eq:4.5.2}) would have been ordered to the right in front of the volume operator $|v|^{\frac{1}{4}}$, then this term would not appear as a correction to (\ref{eq:4.5.10}).\\
Moreover, in \cite{DEath:1996ejf,DEath:1992wve}, in contrast to the present situation, the exponential term appears in the maximally fermion occupied state. This is simply due to the fact that there $\phi_{A'}$ has been implemented as a derivation whereas $\bar{\phi}_A$ and thus $\pi_{\phi}^{A'}$ is quantized as a multiplication operator. Since this choice and the choice made here just correspond to two different representations of the CAR $*$-algebra, they are simply related via a \emph{Bogoliubov transformation}.\\
In fact, in \cite{DEath:1996ejf,DEath:1992wve}, one considers a semi-classical approximation of the Hartle-Hawking wave function 
\begin{equation}
\Psi:=\int_{\mathcal{C}}\mathcal{D}[e^I_{\mu}]\mathcal{D}[\psi^A_{\mu}]\mathcal{D}[\bar{\psi}^{A'}_{\mu}]\,\exp\left(-\frac{1}{\hbar}I\right)
\label{eq:4.5.11}
\end{equation} 
with $I$ the Euclidean action and $\mathcal{C}$ the class of compact 4D metrics and fermionic matter fields satisfiying certain prescribed conditions on a given surface. It is then argued, under certain assumptions on the initial conditions and provided, in the reduced setting, $\phi_{A'}$ is quantized as multiplication operator, a semi-classical approximation of (\ref{eq:4.5.11}) is indeed proportional to (\ref{eq:4.5.10}) in case $D=0$.  
\section{Discussion and outlook}
\label{sec:disc}
In the previous sections we have quantized a class of symmetry reduced models of $\mathcal{N}=1$ supergravity using self-dual variables. We have tried to keep the supersymmetry manifest as far as possible, and used ideas and techniques from loop quantum gravity. In particular:
\begin{itemize}
    \item We reduced the full theory to a homogeneous and quasi-isotropic one and show that the essential part of the constraint algebra in the classical theory closes.
    
    \item We calculated the elementary super-holonomies in the reduced theory and show that they can be obtained from simple building blocks which form a superalgebra. \item We find a representation on a super Hilbert space that satisfies the reality conditions. The residual Gauss constraint can be implemented and selects a sub-Hilbert space. 
    \item The supersymmetry constraints and the Hamiltonian constraint can be implemented. Choosing the ordering appropriately reproduces the relevant part of the (super) Dirac algebra of the classical theory. 
    \item We showed that the zero volume state decouples from the dynamics, hence the classical singularity is resolved in this sense. 
    \item In the limit of sending the area gap to zero, a part of the solution space obtained in another approach is recovered. 
\end{itemize}
We would like to highlight the following points. 

Some of the apparent difficulties with quantizing the full theory while maintaining manifest supersymmetry are the necessity of a non-compact structure group, the complicated reality conditions, the fact that Haar measures on quantum groups are typically not positive definite, and the complicated constraints \cite{Eder:2020erq,Konsti-SUGRA:2020}. However, one can take some encouragement from the fact that in the model we considered, these problems turned out to be solvable. In particular, reality conditions select a suitable measure and lead to a Hilbert space that is very close to the one used in standard LQC. For the bosonic sector this was already observed in \cite{Wilson-Ewing:2015lia}, but here we see that it extends to the fermions as well. Similarly, the closure of the constraint algebra actually reduces the quantization ambiguities and thus is a helpful criterion in the process of writing down the quantum theory. 

We started with a theory with manifest supersymmetry but ended with one where parts of it live on in the constraint algebra, but it is not manifest anymore. It is interesting to see where this change happened. Note that \eqref{eq:4.1.7} still contains (the matrix elements of) a super-holonomy, i.e., an element of the relevant supergroup and covariant under supersymmetry transformations. For simplicity, and to obtain a result that can be compared to standard LQC, we then considered an algebra of ``building blocks'' for the matrix elements. But these lack the fine tuning between odd and even degrees of freedom that is necessary for supersymmetry. In the future it would be interesting to make a different choice here, and use the methods of \cite{Eder:2020erq} to quantize the super holonomies directly on a suitable space of functions on the supergroup. 

Another point that we would like to note is that both orientations of the triad are treated on an equal footing throughout the work. In fact, factors of $\text{sign}(\det(e))$ enter in many places and are important for the consistency of the theory in both sectors. However, as we have discussed in section \ref{SUSY constraint}, the dynamics as we have written it does not mix the two sectors. It appears that parity does not act in a well defined way on $c$ and hence in the theory. We will leave it to future investigation if there is a way to make parity a well defined operation in these models. 

We have seen that the dynamics of the system resolves the classsical singularity in the sense that the zero volume quantum state decouples. We leave it as a task for future investigations to study the full dynamics of the theory including local supersymmetric matter fields as a relational clock. We expect that this would lead to the appearance of bounce cosmologies as in the non-supersymmetric models of LQC \cite{Bojowald:2001xe,Ashtekar:2006wn, Ashtekar:2006rx}. 

Finally, while we have seen that some solutions to the constraints have similar behaviour to those of D'Eath et al.\ \cite{DEath:1996ejf,DEath:1992wve} in a certain limit, there are also profound differences, such as the nature of the states and quantum constraint equations. Moreover, we seem to see the Hartle-Hawking state, but not the wormhole state of D'Eath et al. This shows that the use of chiral variables and the quantization principles of LQG have interesting implications that should be understood better in the future.  

\section*{Acknowledgements}
We thank Norbert Bodendorfer and Thomas Thiemann for helpful comments and discussions on quantizing supergravity with methods from loop quantum gravity, and Jorge Pullin and Lee Smolin for communications at an early stage of this work. We thank an anonymous referee for helpful comments and suggestions on an earlier version of this manuscript. KE thanks the German Academic Scholarship Foundation (Studienstiftung des Deutschen Volkes) for financial support.

\newpage
\appendix
\section{Spinor calculus}\label{Spinor}
We summarize some important formulas concerning gamma matrices in Minkowksi spacetime and their chiral representation as we will need them in the main text. The Clifford algebra $\mathrm{Cl}(\mathbb{R}^{1,3},\eta)$ of four-dimensional Minkowski spacetime is generated by gamma matrices $\gamma_I$, $I\in\{0,\ldots,3\}$, satisfying the Clifford algebra relations
\begin{equation}
    \left\{\gamma_{I},\gamma_{J}\right\}=2\eta_{IJ}
		\label{eq:A1.1}
\end{equation}
where the Minkowski $\eta$ is chosen with signature $\eta=\mathrm{diag}(-+++)$. In the chiral theory, we are working in the \emph{chiral} or \emph{Weyl representation} in which the gamma matrices take the form
\begin{equation}
\gamma_{I}=\begin{pmatrix}
0 & \sigma_{I}\\
\bar{\sigma}_{I} & 0
\end{pmatrix}\quad\text{and}\quad\gamma_{*}=\begin{pmatrix}
\mathds{1} & 0\\
0&-\mathds{1}
\end{pmatrix}
\label{eq:A1.2}
\end{equation}
with $\gamma_*:=i\gamma_0\gamma_1\gamma_2\gamma_3$ the highest rank Clifford algebra element also commonly denoted by $\gamma_5$ in four spacetime dimensions. Here,
\begin{equation}
\sigma_{I}=(-\mathds{1},\sigma_i)\quad\text{and}\quad\bar{\sigma}_{I}=(\mathds{1},\sigma_i)
\label{eq:A1.3}
\end{equation}
with $\sigma_i$, $i=1,2,3$ the ordinary Pauli matrices. With respect to spinorial indices, (\ref{eq:A1.3}) are written as $\sigma_I^{AA'}$ and $\bar{\sigma}_{I A'A}$, respectively. These can be used to map the internal indices $I$ of the co-frame $e^I$ to spinorial indices setting
\begin{equation}
e_{\mu}^{AA'}=e^I_{\mu}\sigma_I^{AA'}
\label{eq:A1.4}
\end{equation}
Primed and unprimed Spinor indices are raised and lowered with respect to the complete antsymmetric symbols $\epsilon^{A'B'}$ and $\epsilon^{AB}$, respectively, with the convention
\begin{equation}
\psi_A=\psi^B\epsilon_{BA}\quad\text{and}\quad\psi^A=\epsilon^{AB}\psi_B
\label{eq:A1.5}
\end{equation}
and analogously for primed indices. Due to $\epsilon\sigma_i\epsilon=\sigma_i^T$, one has the useful formula
\begin{equation}
\sigma_{I AA'}=\sigma_I^{BB'}\epsilon_{BA}\epsilon_{B'A'}=-\bar{\sigma}_{I A'A}
\label{eq:A1.6}
\end{equation}
Using (\ref{eq:A1.3}) as well as (\ref{eq:A1.6}), it is easy to see that
\begin{equation}
\sigma^I_{AA'}\sigma_J^{AA'}=-2\delta_J^I
\label{eq:A1.7}
\end{equation}
Finally, in a $3+1$-decomposition $M\cong\mathbb{R}\times\Sigma$ of the four dimensional spacetime $M$, one considers the spinor-valued one-forms $e_a^{AA'}$ which are related to the spatial metric $q$ on $\Sigma$ according to
\begin{equation}
2q_{ab}=-e_{a AA'}e_b^{AA'}
\end{equation}
with $a=1,2,3$. These, together with the future-directed unit normal vector field $n^{AA'}$ which is normal to the time slices $\Sigma_t$ and satisfies
\begin{equation}
n_{AA'}e_a^{AA'}=0\quad\text{and}\quad n_{AA'}n^{AA'}=2
\label{eq:A1.8}
\end{equation}
form a basis of spinors with one primed and one unprimed index. Finally, let us mention the important identities
\begin{align}
n_{AA'}n^{AB'}&=\delta_{A'}^{B'}\label{eq:A1.9}\\
n_{AA'}n^{BA'}&=\delta_{A}^{B}\label{eq:A1.10}\\
\sigma_{i AA'}\sigma^{AB'}_j&=-\delta_{ij}\delta_{A'}^{B'}-i\tensor{\epsilon}{_{ij}^k}n_{AA'}\sigma^{AB'}_k\label{eq:A1.11}\\
\sigma_{i AA'}\sigma^{BA'}_j&=-\delta_{ij}\delta_{A}^{B}+i\tensor{\epsilon}{_{ij}^k}n_{AA'}\sigma^{BA'}_k
\label{eq:A1.12}
\end{align}

\section{Canonical decomposition of Holst action of \texorpdfstring{$\mathcal{N}=1$, $D=4$}{TEXT} anti-de Sitter supergravity}\label{section:Holst}
In \cite{Konsti-SUGRA:2020}, the canonical analysis of the Holst action of non-extended $D=4$ anti-de Sitter supergravity for arbitrary Barbero-Immirzi parameters was considered. As we will need the results for the special case $\beta=\pm i$ in the main text, especially in section \ref{section:2}, let us briefly state the form of the constraints in the canonical theory.\\
The Lagrangian density of the theory takes the form 
\begin{equation}
\mathcal{L}_{Holst-AdS}=\frac{1}{\kappa\beta}E_i^a\dot{A}^{i_a}-\pi_{\alpha}^a\dot{\psi}_a^{\alpha}-A_t^iG_i-\psi_tS+N^aH_a+NH
\end{equation}
with $A_a^i=\Gamma^i_a+\beta K_a^i$ the Ashtekar-Barbero connection, $E_i^a=\sqrt{q}e_i^a$ the dual electric field, $\psi^{\alpha}_a$ the Rarita-Schwinger field and $\psi^a_{\alpha}$ the canonically conjugate momentum which is related to $\psi^a_{\alpha}$ via the reality condition
\begin{align}
\pi^a=-\tensor{\epsilon}{^{abc}}\bar{\psi}_{b}\gamma_{c}\frac{\mathds{1}+i\beta\gamma_{*}}{2\beta}
\end{align}
The Gauss constraint of the theory takes the form
\begin{equation}
G_i=\frac{1}{\kappa\beta}D_a^{(A)}E^a_i-\frac{i}{2}\pi^{a}\gamma_{*}\gamma_{0i}\psi_a
\end{equation}
with $D^{(A)}_aE_i^a=\partial_aE^a_i+\tensor{\epsilon}{_{ij}^k}A_a^jE_k^a$. The vector and Hamiltonian constraint are given by
\begin{align}
H_a:=\frac{1}{\kappa\beta}E_i^{b}F(A)^i_{ab}-\tensor{\epsilon}{^{bcd}}\bar{\psi}_{b}\gamma_{a}\frac{\mathds{1}+i\beta\gamma_{*}}{2\beta}D^{(A)}_{c}\psi_{d}+\frac{1+\beta^2}{2\beta}\tensor{\epsilon}{^{bcd}}K_{[ca]}\bar{\psi}_{b}\gamma_{0}\psi_{d}
\end{align}
and
\begin{align}
    H=&\frac{E^a_iE^b_j}{2\kappa\sqrt{q}}\tensor{\epsilon}{^{ij}_{k}}\left(\tensor*{{F(A)}}{*^k_{ab}}-(1+\beta^2)\tensor{\epsilon}{^k_{mn}}K^m_a K^n_b\right)+\frac{1}{2L}\sqrt{q}\bar{\psi}_a\gamma^{ab}\psi_b+\frac{3}{\kappa L^2}\sqrt{q}\nonumber\\
    &+\tensor{\epsilon}{^{abc}}\bar{\psi}_{a}\gamma_{0}\frac{\mathds{1}+i\beta\gamma_{*}}{2\beta}D^{(A)}_{b}\psi_{c}+\frac{1+\beta^2}{4\beta}\epsilon^{abc}K_b^i\bar{\psi}_a\gamma_{0i}\psi_c
\end{align}
respectively. Here, $D_a^{(A)}\psi_b=\partial_a\psi_b+\frac{i}{2}A_a^i\gamma_{*}\gamma_{0i}\psi_b$ is the exterior covariant derivative induced by $A_a^i$ in the restricted Majorana representation of the $\mathfrak{su}(2)$ subalgebra of $\mathfrak{spin}^+(1,3)$ generated by $M_{jk}=\frac{1}{2}\gamma_{jk}$. Finally, for the supersymmetry constraint is was found
\begin{align}
S=&\epsilon^{abc}\gamma_a\frac{\mathds{1}+i\beta\gamma_{*}}{2\beta}D^{A}_b\psi_c+\frac{\mathds{1}+i\beta\gamma_{*}}{2\beta}D^A_a\left(\epsilon^{abc}\gamma_b\psi_c\right)\nonumber\\
&-\frac{1+\beta^2}{2\beta}\epsilon^{abc}\gamma_0\psi_cK_{ba}-\frac{1}{L}E^{a}_i\gamma^{0i}\psi_a
\label{eq:B1}
\end{align}
For the discussion in section \ref{section:2}, we derive another form of this constraint rewriting the second term in (\ref{eq:B1}). Using anti commutativity of the fermionic fields, it follows
\begin{align}
\bar{\psi}_t\frac{\mathds{1}+i\beta\gamma_{*}}{2\beta}D^A_a\left(\epsilon^{abc}\gamma_b\psi_c\right)&=\epsilon^{abc}\left[\bar{\psi}_t\frac{\mathds{1}+i\beta\gamma_{*}}{2\beta}\partial_a\left(\gamma_b\psi_c\right)+\frac{i}{2}A_a^i\bar{\psi}_t\frac{\mathds{1}+i\beta\gamma_{*}}{2\beta}\gamma_{*}\gamma_{0i}\gamma_b\psi_c\right]\nonumber\\
&=\epsilon^{abc}\left[-\partial_a\left(\bar{\psi}_c\gamma_b\right)\frac{\mathds{1}+i\beta\gamma_{*}}{2\beta}\psi_t+\frac{i}{2}A_a^i\bar{\psi}_c\gamma_b\gamma_{*}\gamma_{0i}\frac{\mathds{1}+i\beta\gamma_{*}}{2\beta}\psi_t\right]\nonumber\\
&=D_a^A\left(\epsilon^{abc}\bar{\psi}_b\gamma_c\right)\frac{\mathds{1}+i\beta\gamma_{*}}{2\beta}\psi_t\nonumber\\
&=-\left(D_a^A\pi^a\right)\psi_t
\end{align}
with $D_a^A\pi^b=\partial_a\pi^b-\frac{i}{2}\pi^b A_a^i\gamma_*\gamma_{0i}$ the exterior covariant derivative in the dual representation. Hence, 
\begin{align}
\bar{\psi}_t S=&\bar{\psi}_t\epsilon^{abc}\gamma_a\frac{\mathds{1}+i\beta\gamma_{*}}{2\beta}D^{A}_b\psi_c-\bar{\psi}_t\frac{1}{L}E^{a}_i\gamma^{0i}\psi_a-\left(D_a^A\pi^a\right)\psi_t\nonumber\\
&-\frac{1+\beta^2}{2\beta}\epsilon^{abc}\gamma_0\psi_cK_{ba}
\label{eq:B2}
\end{align}


\end{document}